# Metalorganic chemical vapor deposition of β-$(Al_xGa_{1-x})_2O_3$ thin films on (001) β-$Ga_2O_3$ substrates

A F M Anhar Uddin Bhuiyan ; Lingyu Meng ; Hsien-Lien Huang ; Jith Sarker ; Chris Chae; Baishakhi Mazumder ; Jinwoo Hwang ; Hongping Zhao

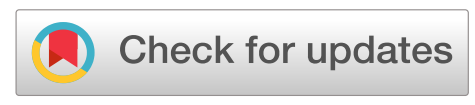

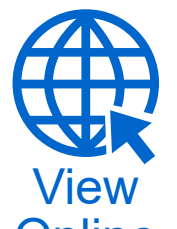

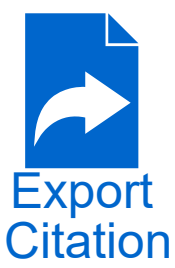

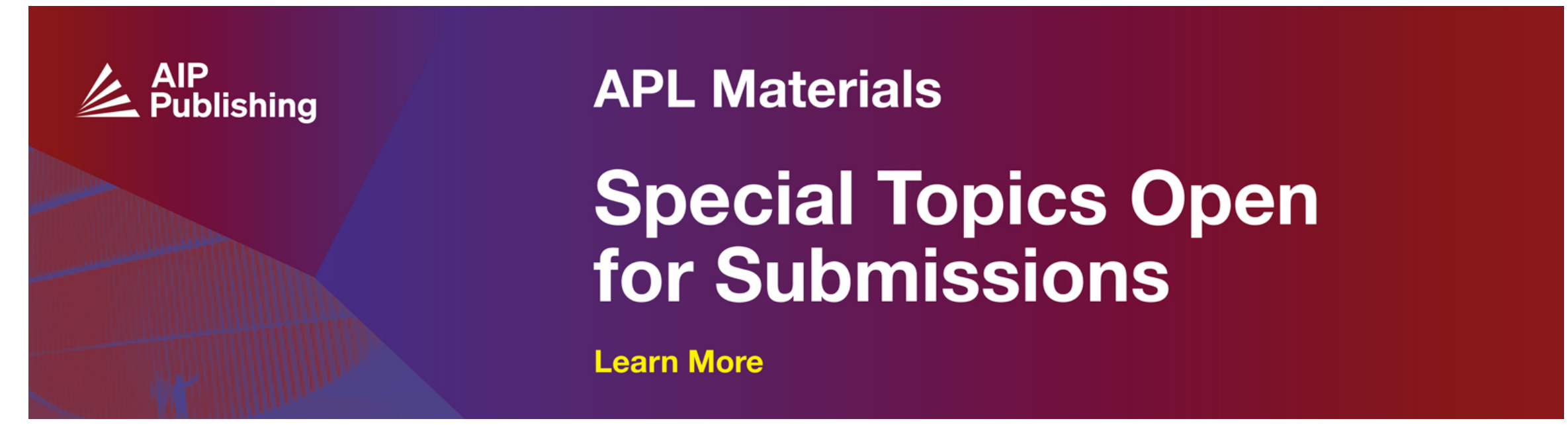

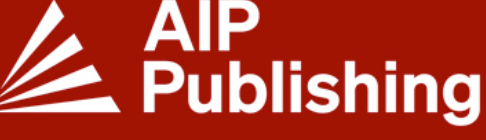

# Metalorganic chemical vapor deposition of β-$(Al_xGa_{1-x})_2O_3$ thin films on (001) β-$Ga_2O_3$ substrates

A F M Anhar Uddin Bhuiyan[1 a)] Lingyu Meng,[1] Hsien-Lien Huang,[2] Jith Sarker,[3] Chris Chae,[2] Baishakhi Mazumder,[3] Jinwoo Hwang,[2] and Hongping Zhao[1,2,b)]

## AFFILIATIONS

[1] Department of Electrical and Computer Engineering, The Ohio State University, Columbus, Ohio 43210, USA
[2] Department of Materials Science and Engineering, The Ohio State University, Columbus, Ohio 43210, USA
[3] Department of Materials Design and Innovation, University at Buffalo, Buffalo, New York 14260, USA

[a)] **E-mail:** bhuiyan.13@osu.edu
[b)] **Author to whom correspondence should be addressed:** zhao.2592@osu.edu

## ABSTRACT

Phase pure β-$(Al_xGa_{1-x})_2O_3$ thin films are grown on (001) oriented β-$Ga_2O_3$ substrates via metalorganic chemical vapor deposition. By systematically tuning the precursor molar flow rates, the epitaxial growth of coherently strained β-$(Al_xGa_{1-x})_2O_3$ films is demonstrated with up to 25% Al compositions as evaluated by high resolution x-ray diffraction. The asymmetrical reciprocal space mapping confirms the growth of coherent β-$(Al_xGa_{1-x})_2O_3$ films (x < 25%) on (001) β-$Ga_2O_3$ substrates. However, the alloy inhomogeneity with local segregation of Al along the $(\bar{2}01)$ plane is observed from atomic resolution STEM imaging, resulting in wavy and inhomogeneous interfaces in the β-$(Al_xGa_{1-x})_2O_3$/β-$Ga_2O_3$ superlattice structure. Room temperature Raman spectra of β-$(Al_xGa_{1-x})_2O_3$ films show similar characteristics peaks as the (001) β-$Ga_2O_3$ substrate without obvious Raman shifts for films with different Al compositions. Atom probe tomography was used to investigate the atomic level structural chemistry with increasing Al content in the β-$(Al_xGa_{1-x})_2O_3$ films. A monotonous increase in chemical heterogeneity is observed from the in-plane Al/Ga distributions, which was further confirmed via statistical frequency distribution analysis. Although the films exhibit alloy fluctuations, n-type doping demonstrates good electrical properties for films with various Al compositions. The determined valence and conduction band offsets at β-$(Al_xGa_{1-x})_2O_3$/β-$Ga_2O_3$ heterojunctions using x-ray photoelectron spectroscopy reveal the formation of type-II (staggered) band alignment.

## I. INTRODUCTION

Aluminum gallium oxide [$(Al_xGa_{1-x})_2O_3$] with its tunable ultrawide bandgap energy (4.87–8.82 eV)[1] and high predicted breakdown field strength ($E_{Br}$ < 16 MV/cm, x < 0.8)[2] has recently been emerged as a promising semiconducting material for next-generation high-power and high-frequency electronic and ultraviolet optoelectronic applications. Considering its capability of controllable n-type doping,[3] β-$(Al_xGa_{1-x})_2O_3$-based vertical power devices can gain benefits in terms of high breakdown field strength with a higher Baliga's figure of merit (BFOM) for a variety of compositions,[2] potentially exceeding that of $Ga_2O_3$. In addition, the formation of a two-dimensional electron gas (2DEG) channel near the β-$(Al_xGa_{1-x})_2O_3$/$Ga_2O_3$ interfaces by modulation doping enhances the carrier density and electron mobility in lateral devices via the screening of the polar optical phonon scattering.[4,5] A higher 2DEG charge can be achieved in the channel with enhanced mobility by increasing the Al content in the β-$(Al_xGa_{1-x})_2O_3$ layer due to the increased conduction band offsets that facilitate better carrier confinement at β-$(Al_xGa_{1-x})_2O_3$/$Ga_2O_3$ heterointerfaces.[6] The recent demonstrations of high-performance modulation doped field-effect transistors (MODFETs) based on the β-$(Al_xGa_{1-x})_2O_3/Ga_2O_3$ heterostructure[5,7,8] have shown great promises of this material in applications such as integrated power and radio-frequency electronics.

Incorporating high Al composition in $(Al_xGa_{1-x})_2O_3$ films is desired to achieve maximized mobility and high carrier density in lateral devices and high breakdown field strength in vertical power devices. However, the maximum Al incorporation in β-$(Al_xGa_{1-x})_2O_3$ films is found to be limited due to the different ground state crystal structures of two parent materials: thermally stable monoclinic β-$Ga_2O_3$ (space group C2/m) and corundum α-$Al_2O_3$ (space group R$\bar{3}$c).[1] Owing to its highly anisotropic crystalline structure, the epitaxial growth of monoclinic β-$(Al_xGa_{1-x})_2O_3$ has been primarily established on (010) oriented β-$Ga_2O_3$ substrates with high quality and excellent electronic transport characteristics.[9–13] N-type doping of (010) β-$(Al_xGa_{1-x})_2O_3$ films has been demonstrated using Si as a dopant for a range of Al compositions.[9,10] However, our previous studies have shown the appearance of γ-phase in metalorganic chemical vapor deposition (MOCVD) grown (010) oriented β-$(Al_xGa_{1-x})_2O_3$ films while targeting for Al compositions above 27%,[13] which impedes the pathway to achieve higher conduction band offsets at the interfaces between β-$(Al_xGa_{1-x})_2O_3$ and β-$Ga_2O_3$ along the (010) orientation. While achieving high Al composition (010) β-$(Al_xGa_{1-x})_2O_3$ films were energetically unfavorable due to the structural phase transformation and domain rotation,[13–17] we utilized other orientations of β-$Ga_2O_3$ substrates, such as (100)[18] and ($\bar{2}$01)[19] for the growth of phase pure β-$(Al_xGa_{1-x})_2O_3$ films by MOCVD growth methods, which exhibited Al incorporations up to <52%. The band offsets determined at the β-$(Al_xGa_{1-x})_2O_3/Ga_2O_3$ interfaces also exhibited a strong orientation dependance.[20,21] The (100) oriented β-$(Al_xGa_{1-x})_2O_3/Ga_2O_3$ heterointerfaces showed relatively larger conduction band offsets than (010) and ($\bar{2}$01) orientations for a wide range of Al compositions.[21] Although the high Al composition of β-$(Al_xGa_{1-x})_2O_3$ films were achieved along (100) and ($\bar{2}$01) orientations, the n-type doping in these films was found to be challenging due to the formation of incoherent boundary defects and alloy inhomogeneity in the films.

While the epitaxial growth of β-$(Al_xGa_{1-x})_2O_3$ has been extensively studied on (010), (100), and ($\bar{2}$01) oriented β-$Ga_2O_3$ substrates using MOCVD growth methods, the MOCVD epitaxial development of β-$(Al_xGa_{1-x})_2O_3$ films on the (001) oriented β-$Ga_2O_3$ substrate is still lacking. The (001) oriented β-$Ga_2O_3$ substrates have been mostly utilized for the homoepitaxial growth of β-$Ga_2O_3$ films via halide vapor-phase epitaxy (HVPE).[22] Majority of the existing vertical β-$Ga_2O_3$ devices based on Schottky barrier,[23–25] p–n heterojunction,[26] or metal–insulator–semiconductor (MIS)[27] diodes have been fabricated on (001) oriented β-$Ga_2O_3$ films. This indicates a great need for the development of high quality β-$(Al_xGa_{1-x})_2O_3$ films on (001) β-$Ga_2O_3$ substrates. Only recently, coherently strained (001) β-$(Al_xGa_{1-x})_2O_3$ films with up to 15% Al content have been demonstrated in molecular beam epitaxy (MBE) via metal oxide-catalyzed epitaxy using Sn as the surfactant for β-$(Al_xGa_{1-x})_2O_3$ film growths.[28] However, the investigation of the solubility limit of (001) β-$(Al_xGa_{1-x})_2O_3$ films grown using MOCVD is still not reported. In addition, the band offset values at the (001) oriented β-$(Al_xGa_{1-x})_2O_3/Ga_2O_3$ interfaces are still lacking.

In this study, we investigated the MOCVD growth of β-$(Al_xGa_{1-x})_2O_3$ films and β-$(Al_xGa_{1-x})_2O_3/Ga_2O_3$ superlattice structures on (001) β-$Ga_2O_3$ substrates. The structural, electrical, chemical, and surface morphological properties of the MOCVD grown (001) β-$(Al_xGa_{1-x})_2O_3$ films are evaluated by comprehensive material characterization, including XRD, high-resolution STEM, energy dispersive x-ray spectroscopy (EDX), XPS, atomic force microscopy (AFM), field emission scanning electron microscopy (FESEM), atom probe tomography (APT), and Raman spectroscopy. Moreover, the evolution of the band offsets at the (001) β-$(Al_xGa_{1-x})_2O_3/Ga_2O_3$ heterointerfaces with the variation of Al compositions is also investigated by using the XPS measurement.

## II. EXPERIMENTAL SECTION

β-$(Al_xGa_{1-x})_2O_3$ thin films were grown on (001) Fe-doped semi-insulating β-$Ga_2O_3$ substrates (purchased from Novel Crystal Technology) by using an Agnitron Agilis MOCVD reactor. Trimethylaluminum (TMAl) and triethylgallium (TEGa) were used as Al and Ga precursors, respectively. Pure $O_2$ gas was used as the O-precursor, and Argon (Ar) was used as the carrier gas. The growth temperature and the chamber pressure were set at 880 °C and 20 Torr, respectively. The TEGa molar flow rate was fixed at 11.95 $\mu$mol min$^{-1}$, while the TMAl molar flow rate was adjusted from 0.15 to 0.91 $\mu$mol min$^{-1}$. The $O_2$ molar flow rate was set at 500 SCCM. The substrates were first *ex situ* cleaned by using solvents, and then, a high-temperature *in situ* cleaning was performed under $O_2$ atmosphere for 5 min at 920 °C prior to the initiation of the epitaxial growth to remove any potential contamination from the substrate surface.

The Al compositions, crystalline structure, film quality, and strain were evaluated by using XRD measurements using a Bruker D8 Discover with Cu Kα radiation x-ray source (λ = 1.5418 Å). The surface morphology and surface roughness were characterized using FESEM (FEI Helios 600) and AFM (Bruker AXS Dimension Icon), respectively. Room temperature Raman spectroscopy was performed using a laser beam of 514 nm (Renishaw–Smiths Detection Combined Raman-IR Microprobe). Aberration-corrected Thermo Fisher Scientific Themis-Z scanning transmission electron microscopy was used to obtain high angle annular dark field (HAADF) STEM images and EDX spectral mapping. Film thicknesses were obtained from the cross-sectional FESEM using coloaded sapphire substrates and STEM HAADF images as well as from STEM-EDX elemental mapping profile of the films. XPS measurements were performed by using the Kratos Axis Ultra x-ray photoelectron spectrometer with a monochromatized Al Kα x-ray source ($E_{photon}$ = 1486.6 eV) to determine the Al compositions and bandgaps of β-$(Al_xGa_{1-x})_2O_3$ thin films. The valence and conduction band offsets were also determined by utilizing XPS with an energy resolution of 0.1 eV. APT specimens were prepared using the standard lift-out and annular milling method.[29] To protect the surface from any potential ion beam damage, a 50 nm thick nickel (Ni) layer was deposited on top of the sample via electron beam evaporation. The APT data acquisitions were conducted using a pulsed laser-assisted CAMECA Local Electrode Atom Probe (LEAP) 5000X HR system. CAMECA's Integrated Visualization and Analysis Software (IVAS 3.8.6) was used for the tip reconstructions and analysis.

## III. RESULTS AND DISCUSSIONS

High-resolution XRD measurements were used to probe the crystalline quality of the MOCVD grown epi-films. Figures 1(a)–1(c) show the XRD ω-2θ scans for (002) reflections of β-$(Al_xGa_{1-x})_2O_3$ thin films grown on the (001) β-$Ga_2O_3$ substrates. The films were grown using [TMAl]/[TEGa+TMAl] molar flow rate ratios of 2.53%, 3.71% and 5.98% with the targeted thicknesses of 200 nm. The high-intensity diffraction peak at 2θ = 31.83° originates from the (002) reflection of the β-$Ga_2O_3$ substrates. With the increase of the [TMAl]/[TEGa+TMAl] molar flow rate ratio from 2.53% to 5.98%, the diffraction peak corresponding to (002) β-$(Al_xGa_{1-x})_2O_3$ shifts toward higher 2θ angles. The increase in the separation between the substrate and epilayer peak positions with the increase of Al composition in the films corresponds to a decrease of (002) plane spacing, which can be attributed to the smaller atomic radius of Al than Ga.[30] As the Al composition increases, the higher lattice mismatch between the substrate and the films results in higher tensile strain in the epilayer due to the contraction of the lattice constant, causing the peak corresponding to epi-film shifts toward higher 2θ angles.[9,12,31,32] The Al content in β-$(Al_xGa_{1-x})_2O_3$ films are determined by calculating the interplanar distance between the β-$Ga_2O_3$ substrate and β-$(Al_xGa_{1-x})_2O_3$ epi-film from the corresponding XRD peak positions. The Al compositions of 11%, 14%, and 25% are determined for [TMAl]/[TEGa+TMAl] molar flow rate ratios of 2.53%, 3.71%, and 5.98%, respectively. The Al compositions estimated from the interplanar distances between the substrate and epilayers are found to be in good agreement with the Al compositions estimated by assuming fully strained thin films growths on the β-$Ga_2O_3$ substrate.[28] While we observed a monotonous shift of the peak positions of β-$(Al_xGa_{1-x})_2O_3$ epi-films as we increased Al molar flow rates, the Al compositions determined from these peak positions agree well with XPS, STEM-EDX elemental mapping, and APT measurements. As the Al composition increases, the β-$(Al_xGa_{1-x})_2O_3$ peak intensity reduces, indicating the degradation of crystalline quality of the films with higher Al compositions. The low intensity peak of higher Al composition samples [i.e., the 25% Al composition as shown in Fig. 1(c)] can be attributed to its inhomogeneous Al distribution in the film (discussed later using STEM and APT measurements). However, the lower x-ray scattering factor of Al than that of Ga atom may also partially contribute to the

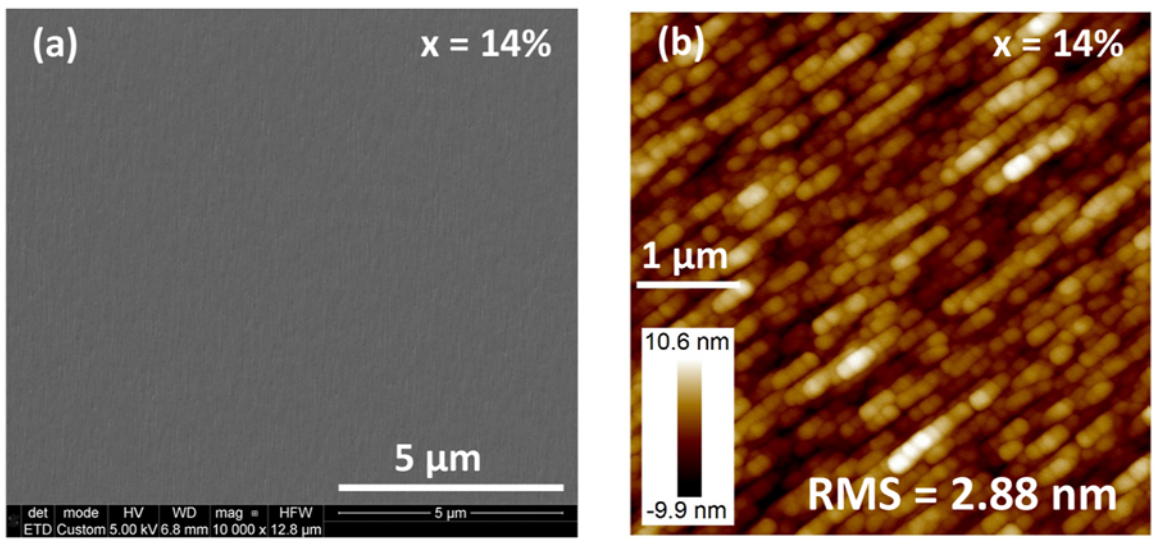

**FIG. 2.** (a) Surface view FESEM and (b) AFM images of β-$(Al_xGa_{1-x})_2O_3$ films with 14% Al compositions.

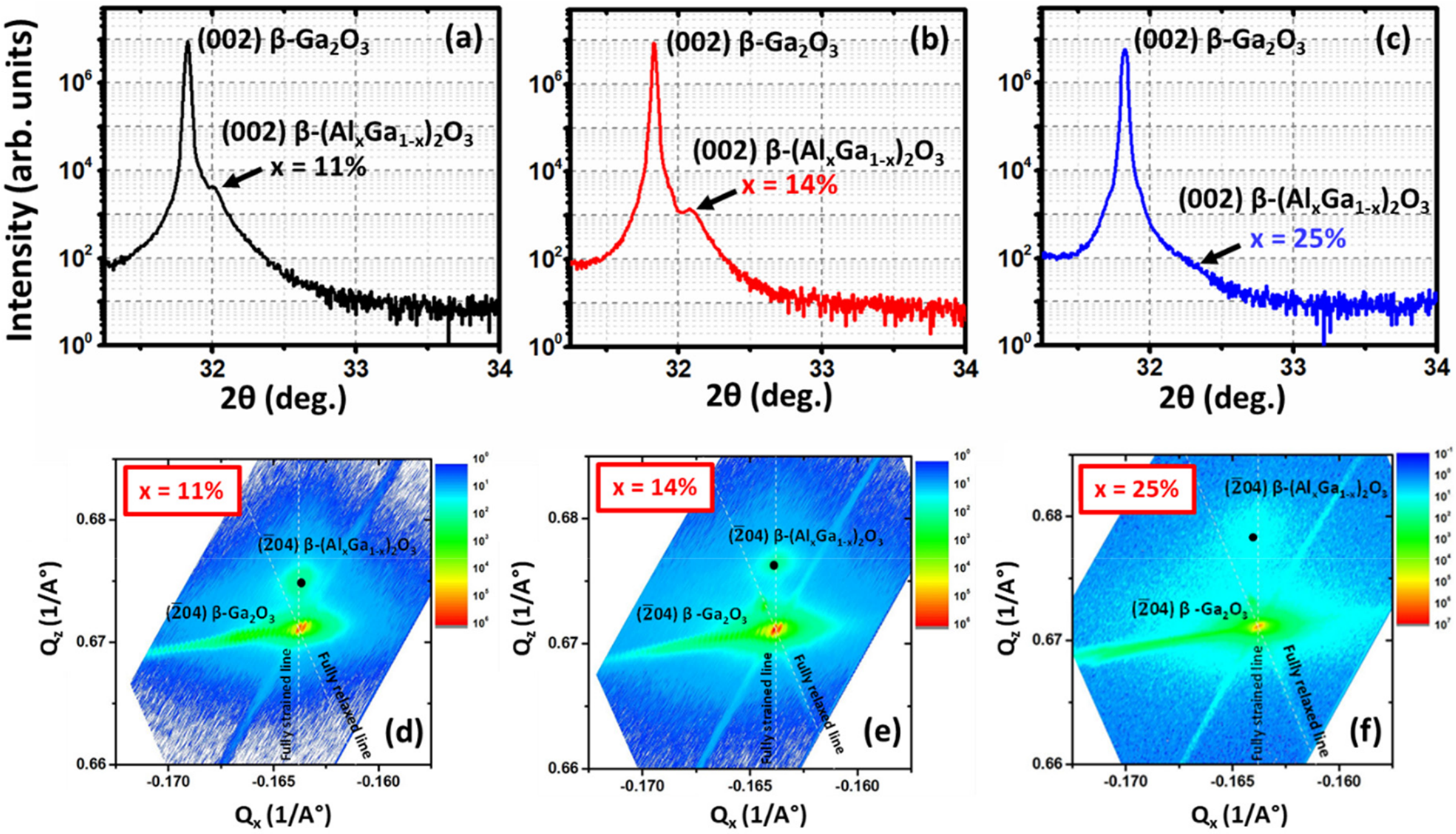

**FIG. 1.** XRD ω-2θ scan profiles of the (002) reflections of β-$(Al_xGa_{1-x})_2O_3$ films grown on (001) β-$Ga_2O_3$ substrates with Al incorporations of (a) 11%, (b) 14%, and (c) 25%. Asymmetrical reciprocal space maps (RSMs) around ($\bar{2}$04) reflections of (001) β-$(Al_xGa_{1-x})_2O_3$ films with (d) x = 11%, (e) x = 14%, and (f) x = 25%. The vertical and tilted white dashed lines represent fully strained and fully relaxed positions, respectively.

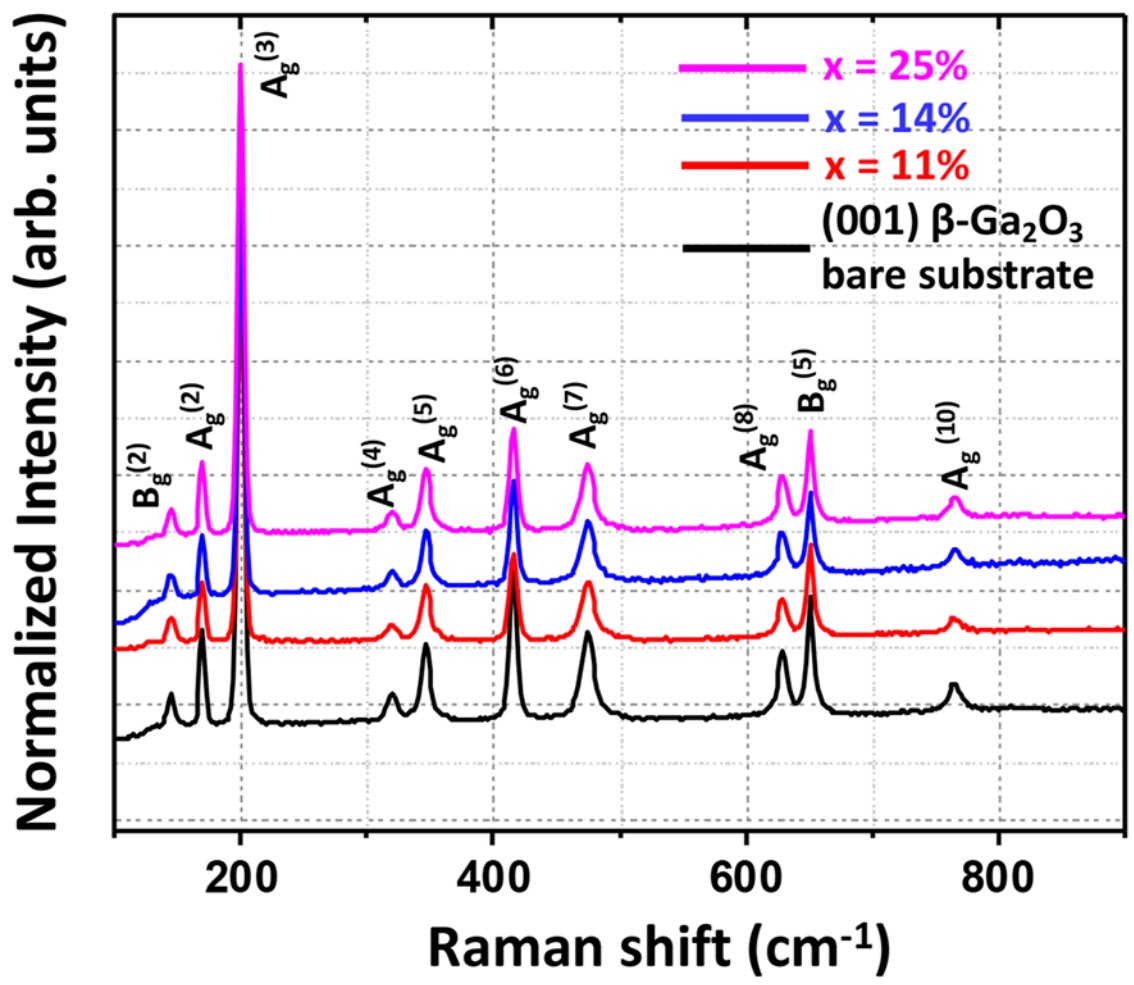

**FIG. 3.** Room-temperature Raman spectra of β-$(Al_xGa_{1-x})_2O_3$ films with Al compositions of 11%, 14%, and 25%. The Raman spectra from the (001) β-$Ga_2O_3$ bare substrate are also included.

reduction of the peak intensity of β-$(Al_xGa_{1-x})_2O_3$ films with higher Al contents.[33]

To investigate the strain state and confirm the coherent growth of (001) β-$(Al_xGa_{1-x})_2O_3$ films on β-$Ga_2O_3$ substrates, the asymmetrical reciprocal space mapping (RSM) was performed on β-$(Al_xGa_{1-x})_2O_3$ films for different Al compositions. Figures 1(d)–1(f) show the asymmetrical RSMs for $(\bar{2}04)$ reflections of β-$(Al_xGa_{1-x})_2O_3$ films with 11%, 14%, and 25% Al compositions, respectively. The fully relaxed and strained positions are shown by the tilted and vertical white dashed lines, respectively. The maximum reflection intensity of the $(\bar{2}04)$ β-$(Al_xGa_{1-x})_2O_3$ reciprocal lattice point moves far from the $(\bar{2}04)$ β-$Ga_2O_3$ substrate peak as the Al composition increases along the fully strained vertical line, indicating the growth of fully coherent β-$(Al_xGa_{1-x})_2O_3$ films on (001) β-$Ga_2O_3$ substrates for different Al compositions. The shifting of the peak position of $(\bar{2}04)$ β-$(Al_xGa_{1-x})_2O_3$ toward higher $Q_z$ values (out-of-plane reciprocal space lattice constant) due to higher Al incorporation indicates the increase of the lattice mismatch between the strained epi-films and substrates. The surface morphology and roughness of β-$(Al_xGa_{1-x})_2O_3$ films for a representative sample with Al composition of 14% are shown in Fig. 2. Uniform surface morphologies with rms roughness of 2.88 nm are measured from the FESEM and AFM images as shown in Figs. 2(a) and 2(b), respectively.

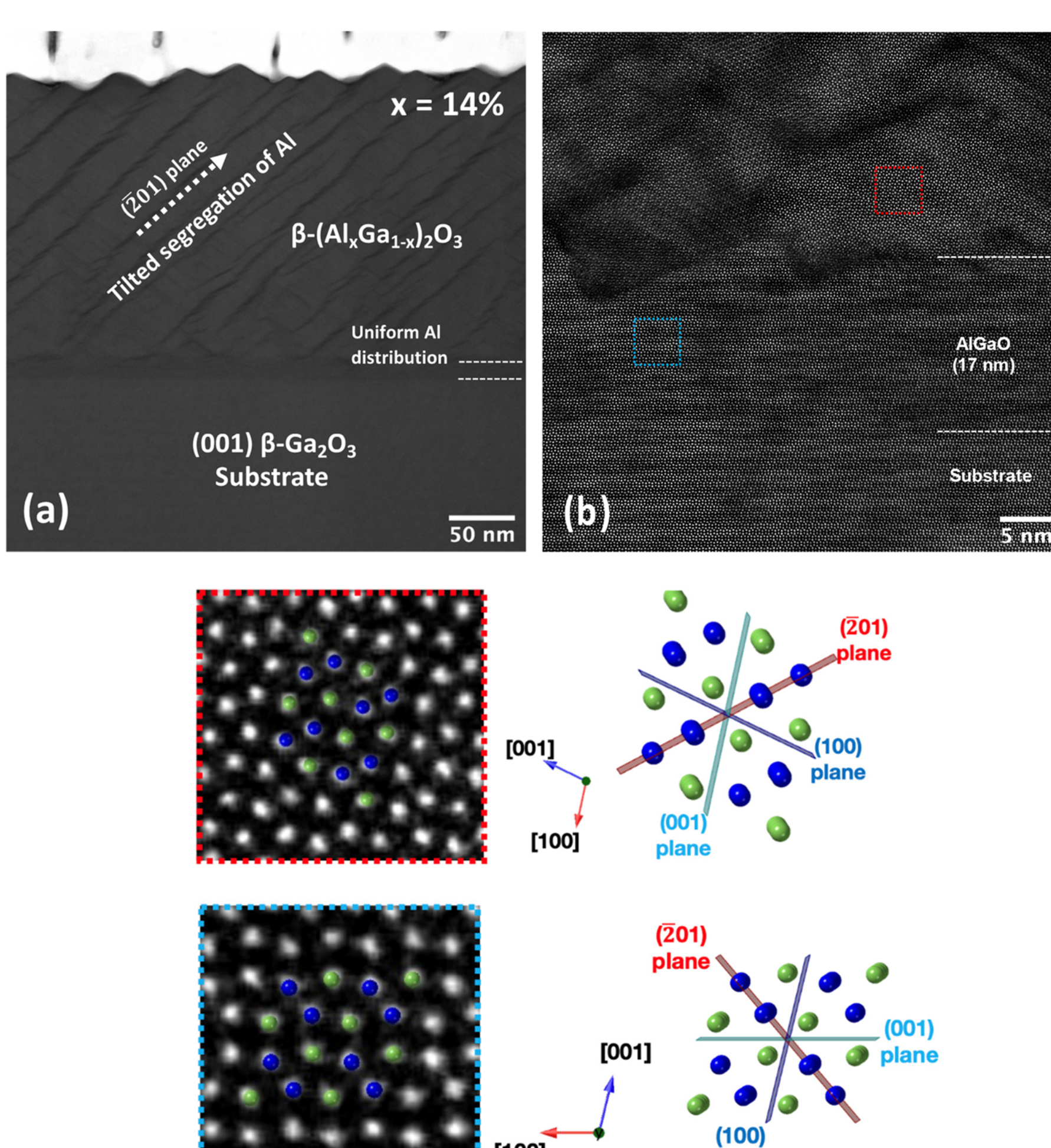

**FIG. 4.** Atomic resolution cross-sectional HAADF-STEM images of β-$(Al_xGa_{1-x})_2O_3$ film with 14% Al compositions at (a) 50 nm and (b) 5 nm scales, showing Al distribution pattern along the $(\bar{2}01)$ cleavage plane with a ~17 nm thick homogeneous and uniform β-$(Al_xGa_{1-x})_2O_3$ layer at the interface. The HAADF-STEM images were taken from the ⟨010⟩ zone axes.

Room temperature Raman spectra of the β-$(Al_xGa_{1-x})_2O_3$ films with different Al compositions are shown in Fig. 3. The Raman spectrum of a single crystal (001) β-$Ga_2O_3$ bare substrate is also included in the figure as reference. In general, the irreducible representation for acoustical and optical zone center modes based on the factor group analysis[34] at the Γ point is $\Gamma_{aco} = A_u + 2B_u$ and $\Gamma_{opt} = 10A_g + 5B_g + 4A_u + 8B_u$, respectively. For the optical modes, $A_g$ and $B_g$ are Raman active, while $A_u$ and $B_u$ modes are infrared active. Both $A_g$ and $B_g$ mode peaks are selectively observed in the Raman spectra of the (001) β-$(Al_xGa_{1-x})_2O_3$ films with different Al compositions, as shown in Fig. 3. Ten Raman peaks are observed from both (001) β-$Ga_2O_3$ bare substrate and β-$(Al_xGa_{1-x})_2O_3$ films. Our experimental Raman mode frequencies correspond well with those reported in the literature based on experimental and theoretical studies on (001) β-$Ga_2O_3$.[35,36] The peaks at around 169.6, 199.49, 319.8, 345.7, 416.2, 474.3, 629.4, and 765.5 $cm^{-1}$ belong to the $A_g$ vibrational mode, and the peaks at around 144.9 and 651.2 $cm^{-1}$ can be assigned to the $B_g$ mode. These Raman-active modes can be classified into three groups: the high-frequency (770–500 $cm^{-1}$) stretching and bending of $GaO_4$ tetrahedra ($A_g^{(8)}$, $B_g^{(5)}$ and $A_g^{(10)}$), the mid-frequency (480–310 $cm^{-1}$) deformation of $Ga_2O_6$ octahedra ($A_g^{(4)}$–$A_g^{(7)}$), and the low-frequency (below 200 $cm^{-1}$) libration and translation of tetrahedra–octahedra chains ($B_g^{(2)}$, $A_g^{(2)}$, and $A_g^{(3)}$).[37] The Raman spectra of the β-$(Al_xGa_{1-x})_2O_3$ films exhibit similar characteristics of the spectrum of the bare β-$Ga_2O_3$ substrate, without any noticeable Raman shifts for different Al compositions, indicating that the symmetry of the β-$(Al_xGa_{1-x})_2O_3$ crystal structures is well-maintained.

In order to investigate the crystalline properties of the materials grown on (001) β-$Ga_2O_3$ substrates, β-$(Al_xGa_{1-x})_2O_3$ thin film and β-$(Al_xGa_{1-x})_2O_3/Ga_2O_3$ superlattice (SL) structure are characterized using high-resolution STEM imaging. Figures 4(a) and 4(b) show the cross-sectional HAADF-STEM images of the β-$(Al_xGa_{1-x})_2O_3$ film with a targeted 14% Al composition. Undisturbed monoclinic β-phase structures without noticeable phase transformation are observed from the STEM images. The sharp contrasts between the (001) β-$Ga_2O_3$ substrate (bright) and β-$(Al_xGa_{1-x})_2O_3$ epi-films (dark) indicate high-quality interface. At the interface, the film displays a ~17 nm thick uniform cross section of the β-$(Al_xGa_{1-x})_2O_3$ layer with homogeneous Al distribution. However, as the growth continues, the compositional segregation with domain rotations is observed, as indicated by the strong tilted contrast in the epitaxial layer as shown in Fig. 4(a). This significant contrast can also be seen at the atomic resolution image in Fig. 4(b), where the non-uniform Al distribution causes fluctuations in atomic column intensity. The Al distribution patterns as indicated by the tilted darker contrast are found to align along the $(\bar{2}01)$ plane, which belongs to the crystal planes having the relatively low surface free energies. Our previous growth studies have shown high-Al incorporation (up to 48%) in β-$(Al_xGa_{1-x})_2O_3$ films grown along $(\bar{2}01)$ orientation,[19] indicating the strong preference of Al incorporation along the $(\bar{2}01)$ plane, which can be related to its significantly lower surface free energy.[38] Such strong anisotropic characteristics of the β-gallia structure leads to the directional dependence of the Al distribution along the $(\bar{2}01)$ plane in β-$(Al_xGa_{1-x})_2O_3$ thin films grown on (001) β-$(Al_xGa_{1-x})_2O_3$ substrates.

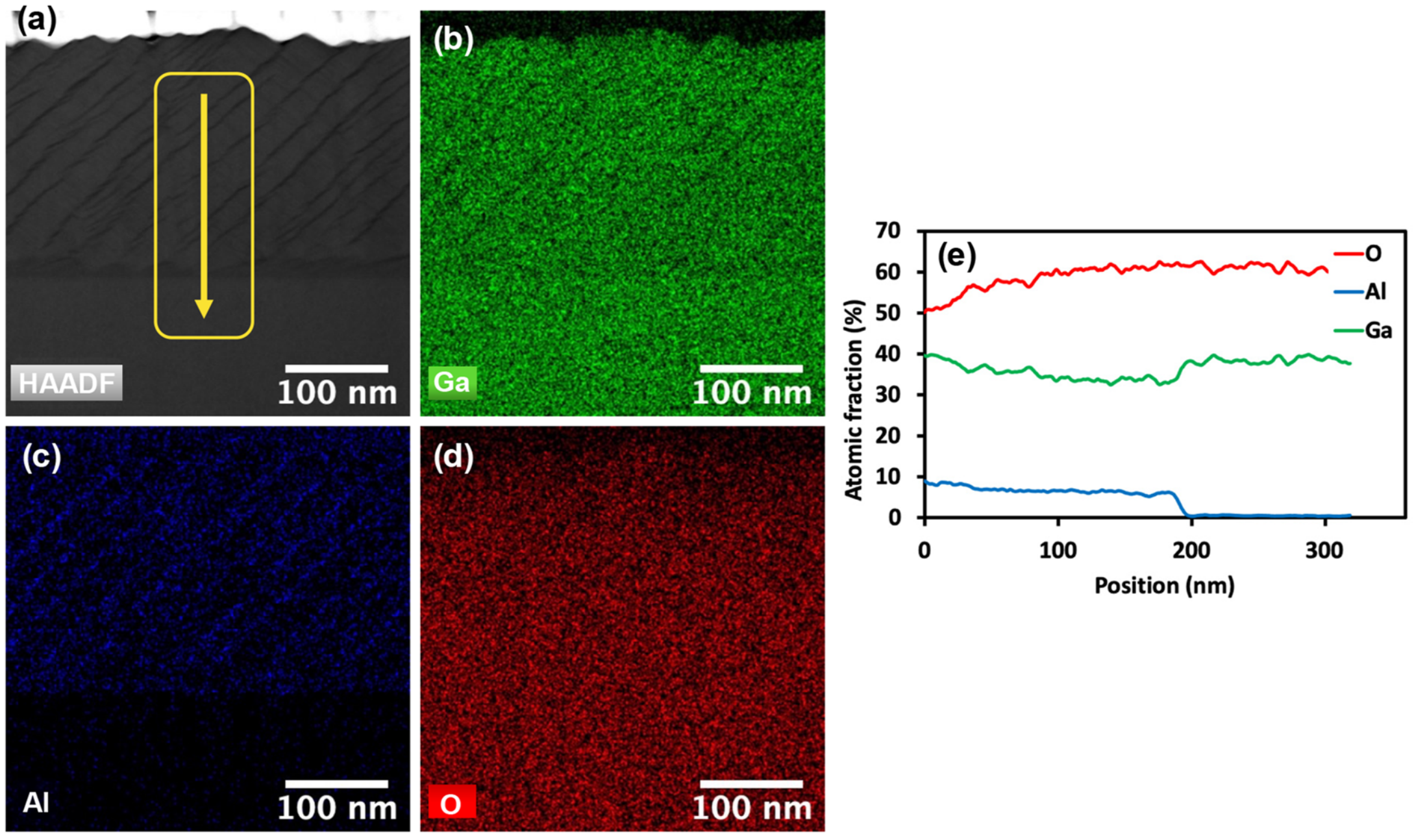

**FIG. 5.** STEM-EDX mapping of β-$(Al_xGa_{1-x})_2O_3$ film with 14% Al compositions. (a) Cross-sectional HAADF image with the corresponding EDX mapping of (b) Ga, (c) Al, and (d) O atoms. (e) Atomic fraction elemental profile as indicated by the orange arrow in (a).

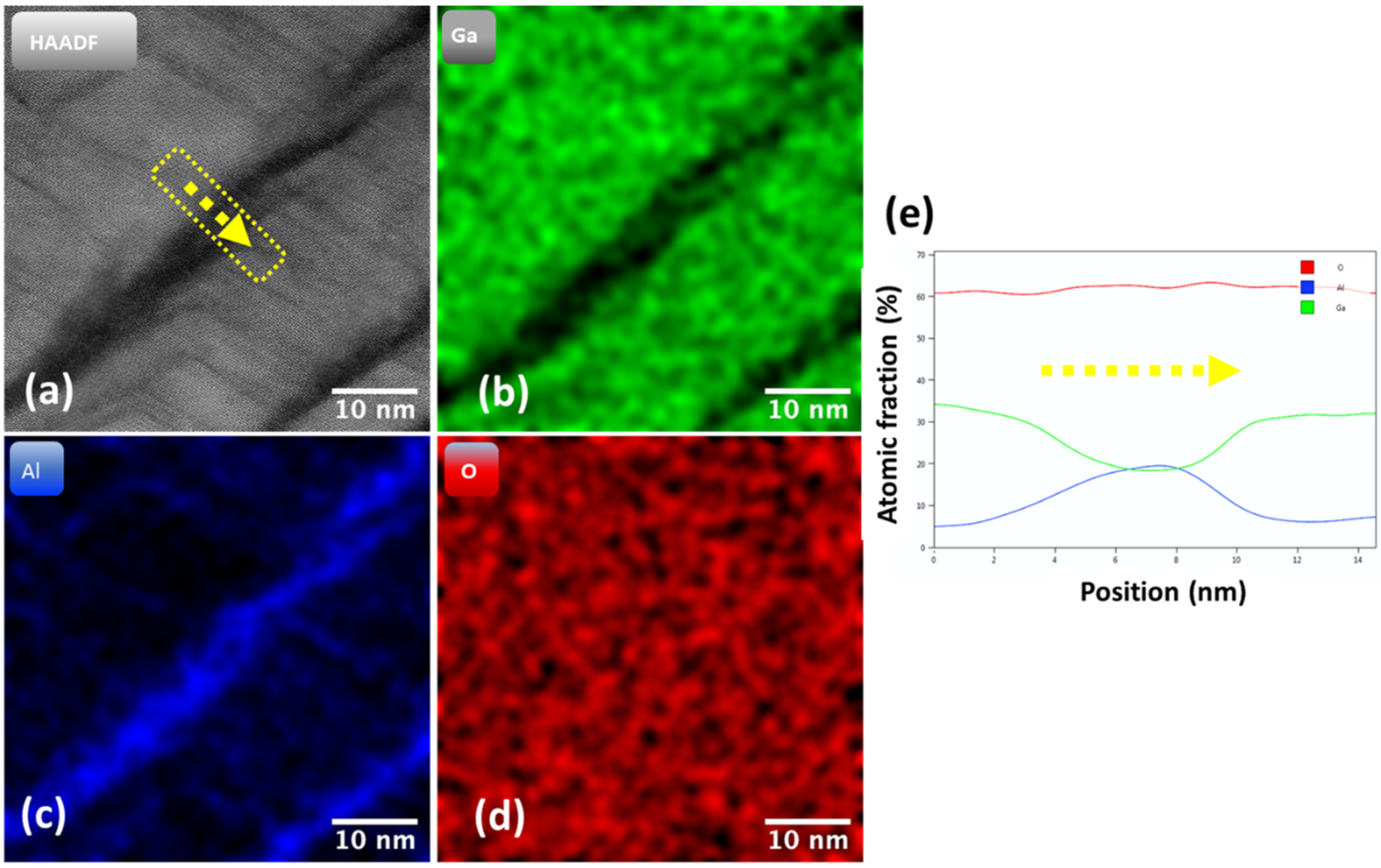

**FIG. 6.** STEM-EDX mapping of β-$(Al_xGa_{1-x})_2O_3$ film with 14% Al compositions, showing a tilted Al distribution region. (a) Cross-sectional HAADF image with the corresponding EDX mapping of (b) Ga, (c) Al, and (d) O atoms. (e) Atomic fraction elemental profile as indicated by the orange arrow in (a).

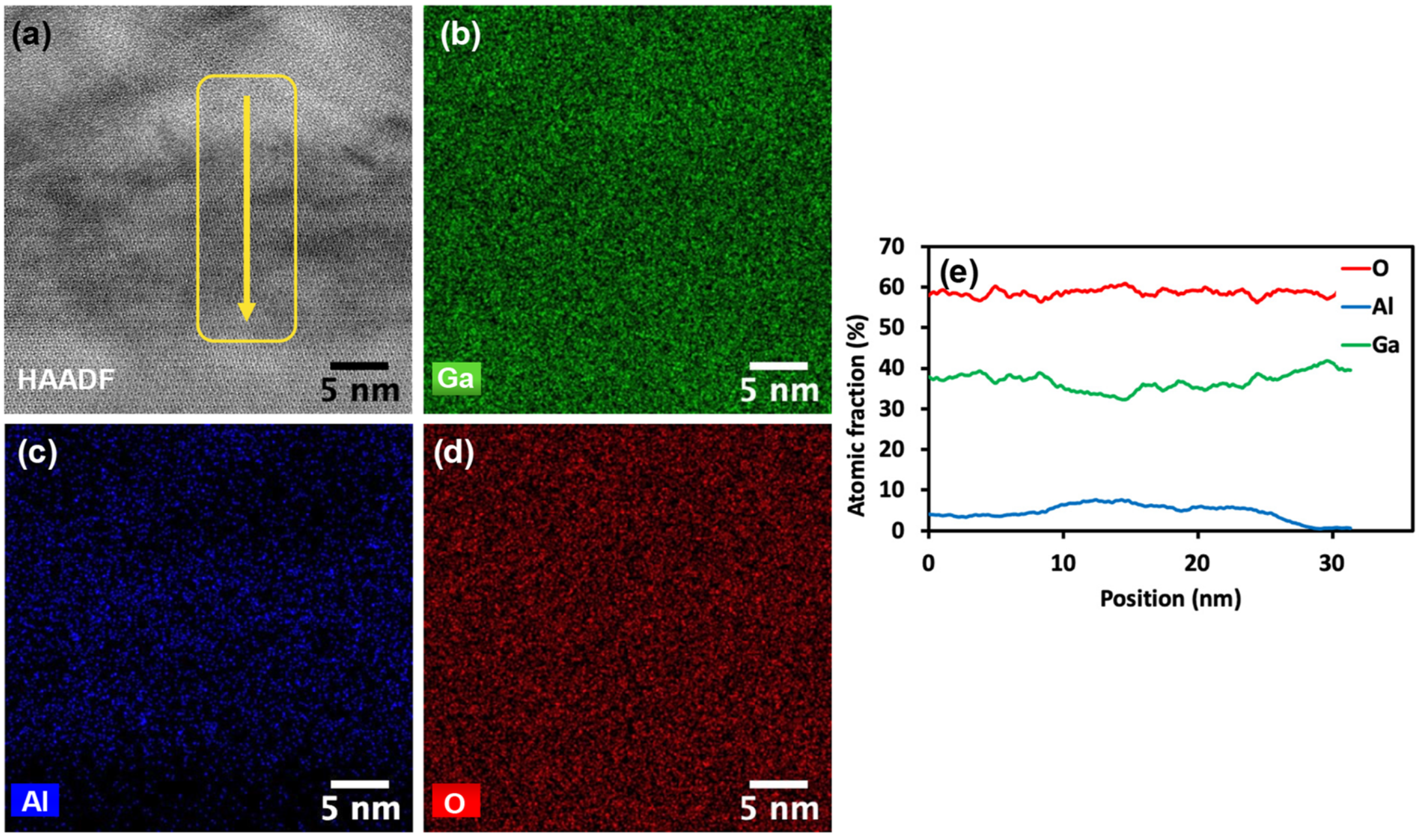

**FIG. 7.** STEM-EDX mapping of β-$(Al_xGa_{1-x})_2O_3$ film with 14% Al compositions, showing a uniform Al distribution region at the interface. (a) Cross-sectional HAADF image with corresponding EDX mapping of (b) Ga, (c) Al, and (d) O atoms. (e) Atomic fraction elemental profile along the orange arrow in (a).

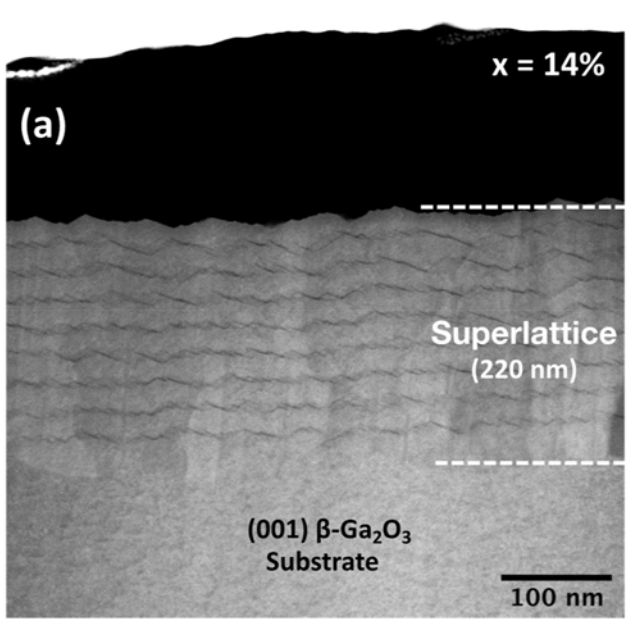

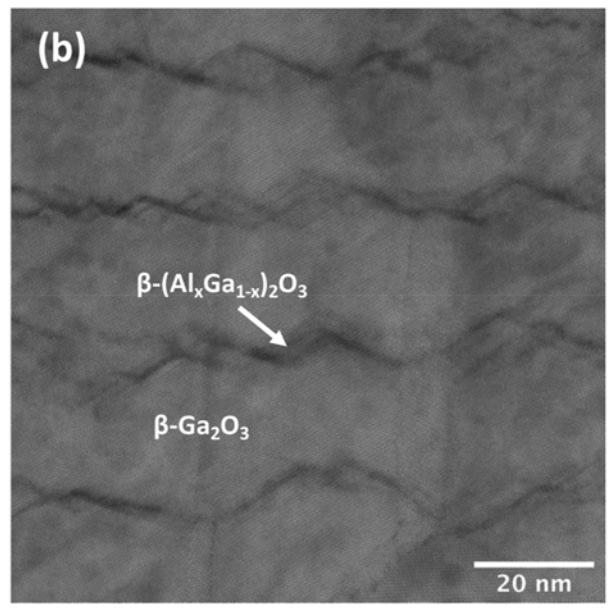

**FIG. 8.** High-resolution cross-sectional HAADF-STEM images of eight period (001) β-$(Al_xGa_{1-x})_2O_3$/β-$Ga_2O_3$ superlattice structure grown with targeted Al composition of 14% in β-$(Al_xGa_{1-x})_2O_3$ layer at (a) 100 nm and (b) 20 nm scales, showing wavy and inhomogeneous interfaces between β-$(Al_xGa_{1-x})_2O_3$ and β-$Ga_2O_3$ layers. The HAADF-STEM images were taken from the ⟨010⟩ zone axes.

Additionally, in order to evaluate the compositional homogeneity and Al composition, STEM-EDX mapping was performed throughout the β-$(Al_xGa_{1-x})_2O_3$ layer grown with 14% Al composition as shown in Fig. 5. The EDX color maps of Ga (green) and Al (blue) elements in Figs. 5(b) and 5(c) reveal the compositional segregation of the films, as also observed from the HAADF STEM images in Fig. 4. The Al distribution pattern in the film along the $(\bar{2}01)$ plane is also confirmed from the EDX color maps of Ga and Al atoms. The average Al compositions (x = ∼16.4%) estimated from the STEM-EDX elemental maps in Fig. 5(e) matches well with those extracted from XRD measurements (x = 14%). Slightly higher Al compositions extracted from EDX elemental mapping can be due to the nonuniformity of the Al distribution in the films. In order to quantitatively evaluate the Al composition in the film along the $(\bar{2}01)$ plane, we also performed the STEM-EDX mapping, focusing on the region that exhibits inhomogeneous Al distribution as shown in Figs. 6(a)–6(e). The line scan as indicated by the orange arrow in Fig. 6(a) provides the concentration of each element, as shown in Fig. 6(e). The quantitative EDX elemental mapping in Fig. 6(e) confirms Al segregation along the $(\bar{2}01)$ plane with high Al content of 50% as indicated by the tilted darker contrast in the HAADF STEM images in Fig. 6(a). However, Al composition of ∼13% is extracted from the elemental mapping in Fig. 6(e) from other regions of the films that exhibit compositional homogeneity and uniformity. While the compositional segregation is observed in the films due to the Al incorporation tendency along the $(\bar{2}01)$ plane, the first ∼17 nm thick layer at the interfaces reveal homogeneous Al distribution with good uniformity as shown in the darker contrast in the cross-sectional HAADF-STEM image in Fig. 7(a) and STEM-EDX elemental maps in Figs. 7(b)–7(e).

In addition to the growth of thin films, we also investigated the growth of β-$(Al_xGa_{1-x})_2O_3$/β-$Ga_2O_3$ superlattice structure grown with targeted Al compositions of 14% in the β-$(Al_xGa_{1-x})_2O_3$ barrier layer. Figures 8(a) and 8(b) show the cross-sectional HAADF-STEM images for the SL structure at 100 and 20 nm scales, respectively. The SL structure is grown on a 65 nm thick β-$Ga_2O_3$ buffer layer. The targeted thickness of the β-$(Al_xGa_{1-x})_2O_3$ barrier and β-$Ga_2O_3$ well layers were 5 and 15 nm, respectively. The darker contrast represents the β-$(Al_xGa_{1-x})_2O_3$ barrier, while the brighter contrast corresponds to the $\beta$-$Ga_2O_3$ well layer. The growth of eight periods of alternating SLs, maintaining the β-phase throughout the entire structure, is confirmed by the STEM images. However, due to the nonuniformity of the Al distribution in the film, $\beta$-$(Al_xGa_{1-x})_2O_3$ layers show zigzag shapes with wavy and inhomogeneous interfaces between $\beta$-$(Al_xGa_{1-x})_2O_3$ and $\beta$-$Ga_2O_3$ layers throughout the SL structure. The similar nonuniform structures are also observed from the STEM-EDX maps of the corresponding SL structures as shown in Figs. 9(a)–9(e). The quantitative elemental mapping in Fig. 9(e) shows alternating compositional profiles of the periodic structures with around 12% Al incorporation in the $\beta$-$(Al_xGa_{1-x})_2O_3$ layers.

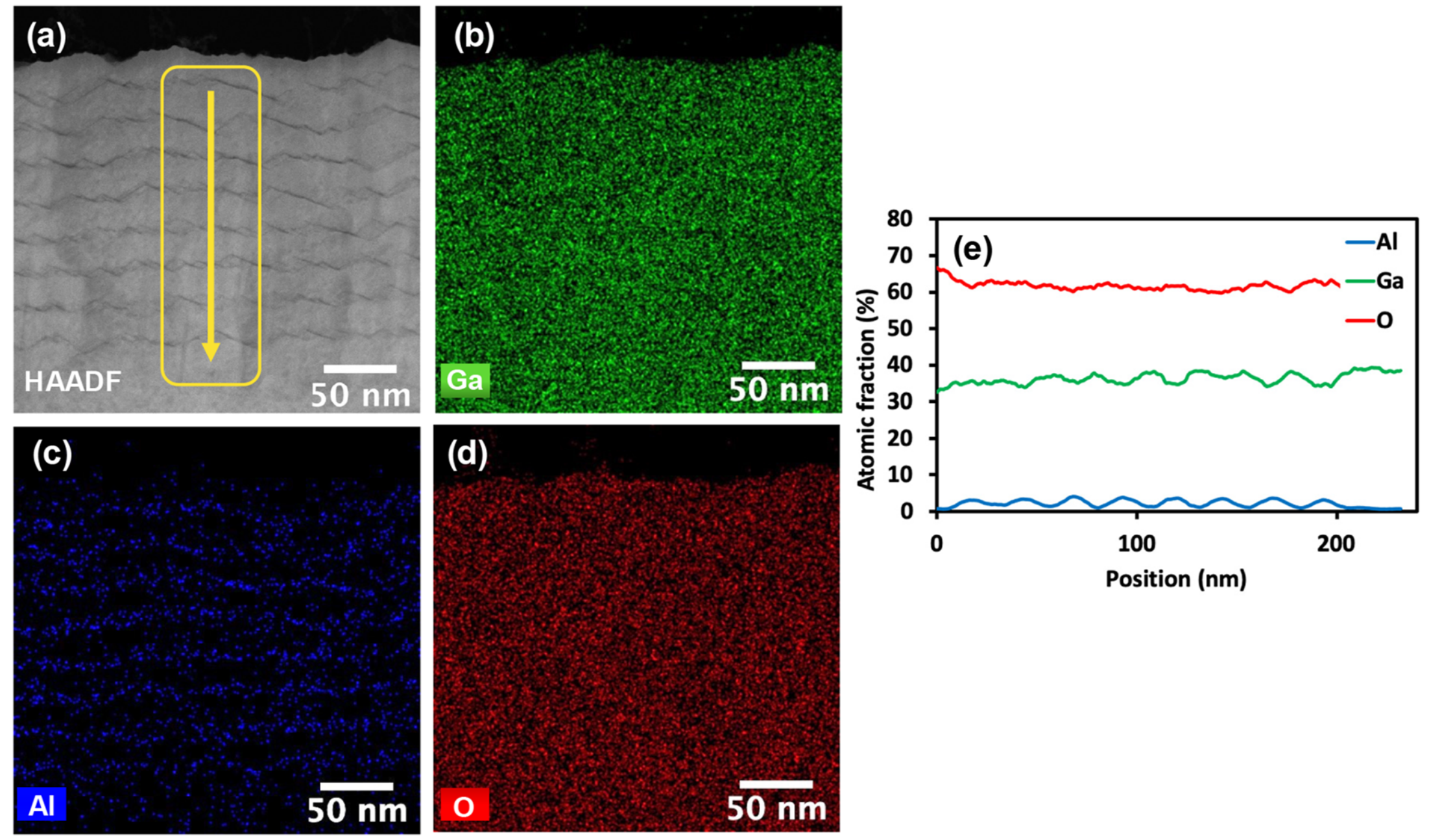

**FIG. 9.** STEM-EDX mapping of (001) β-$(Al_xGa_{1-x})_2O_3$/β-$Ga_2O_3$ superlattice structure grown with targeted Al compositions of 14%. (a) Cross-sectional HAADF image with corresponding EDX mapping of (b) Ga, (c) Al, and (d) O atoms. (e) Atomic fraction elemental profile along the orange arrow in (a), showing eight periods of the SL structure.

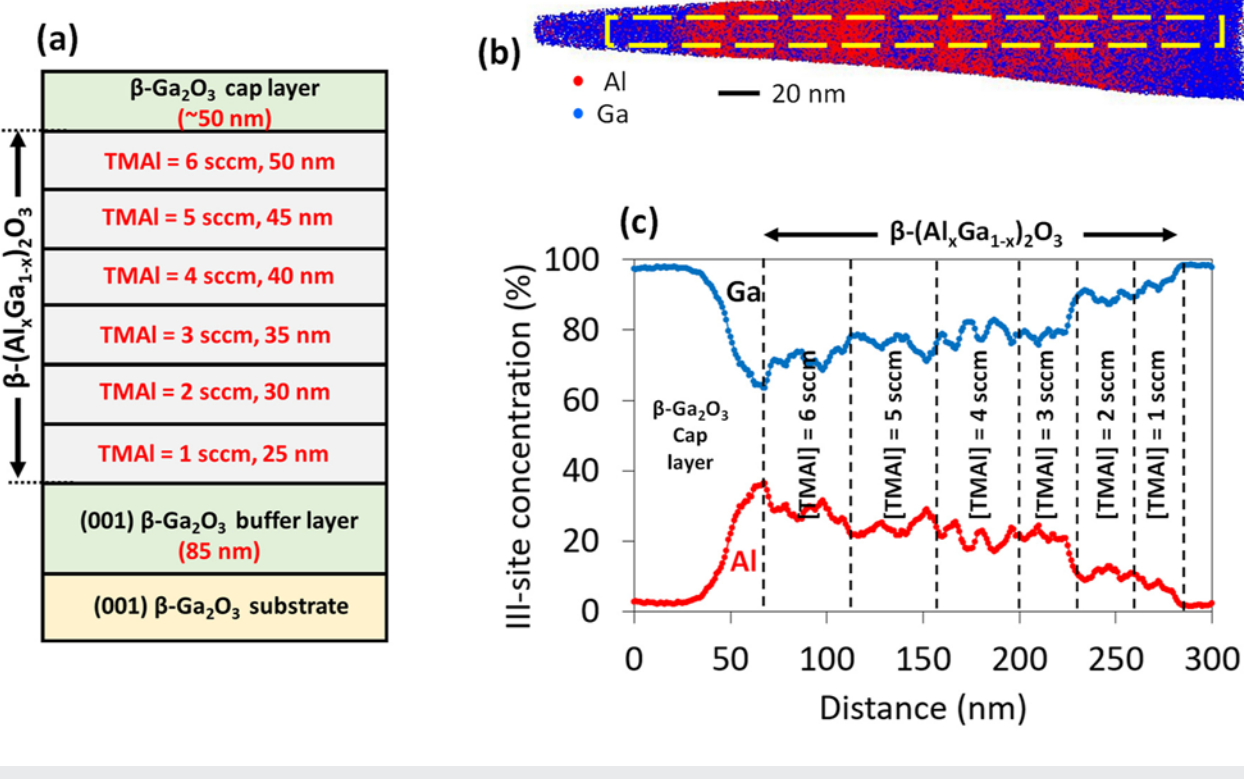

FIG. 10. (a) Schematic of (001) $\beta$-$(Al_xGa_{1-x})_2O_3$/$\beta$-$Ga_2O_3$ heterostructures with varying TMAl flow rates analyzed in APT. (b) A 3D atom map of the corresponding structure—only Al and Ga atoms are shown for clarity in red and blue, respectively. (c) Elemental (III-site) composition profile of the analyzed heterostructure in (a).

APT was used to investigate the atomic scale chemical homogeneity in $\beta$-$(Al_xGa_{1-x})_2O_3$ layers with different Al contents. Figure 10 shows the 3D elemental distribution of a $\beta$-$(Al_xGa_{1-x})_2O_3$ stack grown by varying the TMAl flow rate from 1 to 6 SCCM as shown in the schematic in Fig. 10(a). The red and blue dots correspond to the Al and Ga atoms, respectively. As growth continues, the density of the red dots increases with the increase of the TMAl flow, indicating an increase in the Al contents. Our previous studies on the MOCVD epitaxial growth of (010) oriented $\beta$-$(Al_xGa_{1-x})_2O_3$ thin films have shown excellent alloy homogeneity for the layers grown with relatively low Al compositions ($x < 27\%$), while chemical segregation was observed due to higher Al incorporation.[13] However, in the case of the epitaxial growth of (001) oriented $\beta$-$(Al_xGa_{1-x})_2O_3$ thin films, we observed nonuniformity in the Al distribution even for low Al content layers as shown in the elemental mapping in Fig. 10(c), implying a strong dependence of the chemical inhomogeneity in $\beta$-$(Al_xGa_{1-x})_2O_3$ on the orientations of $\beta$-$Ga_2O_3$ substrates. While nonuniformity in the Al distribution is observed, the average Al compositions from APT are found to be consistent with the values determined from XRD and STEM-EDX mapping.

Additionally, in order to investigate the fluctuation in alloy homogeneity with the increasing Al content in the layer, APT was also utilized to perform lateral Al/Ga distribution and frequency distribution analysis (FDA) of each corresponding layers as shown in the schematic in Fig. 10(a). Figures 11(a)–11(f) show the in-plane Al/Ga ratio for different layers of $\beta$-$(Al_xGa_{1-x})_2O_3$ with increasing Al contents. From each layer, an analysis volume with a diameter of 80 nm and a thickness of 4 nm was chosen. With the increase of the TMAl flow rate from 1 to 6 SCCM, the Al content in the layers increases as indicated by the increase of the Al/Ga ratio. The compositional inhomogeneity increases as the Al composition increases. The higher mobility of Ga adatoms on the growth surface leads to the lower compositional variation in lower Al content layers.[39] However, a strong chemical heterogeneity is observed in the layer grown with higher TMAl flow rates, which can be due to the strong oxidization of Al adatoms[40] lowering the mobility of Al on growth surface. This compositional fluctuation in (001) $\beta$-$(Al_xGa_{1-x})_2O_3$ film can also be correlated with the STEM images as shown in Fig. 5.

The Al distributions within each sub-layers with various TMAl flow rates were further evaluated by statistical FDA, as shown in Figs. 12(a)–12(f), in order to statistically investigate the inhomogeneity in different layers with different Al contents. In general, the chemical inhomogeneity in a layer is indicated by the deviation of the observed elemental distribution from that of a random binomial fitting. The value of the Pearson coefficient, $\mu$ approaching toward 1 indicates statistically significant amount of elemental segregation, and the lower P-value provides a higher confidence level during the null hypothesis testing.[41] For the lower TMAl flow rates, the observed Al distribution follows the expected binomial distribution, implying random distribution of Al atoms. However, the deviation between the observed Al distribution and random binomial fit increases with a higher $\mu$ value, as the TMAl flow rates increase. This phenomenon indicates an increase in chemical inhomogeneity with the increase of the Al content in the films as also observed from the in-plane Al/Ga lateral chemistry maps in Fig. 11.

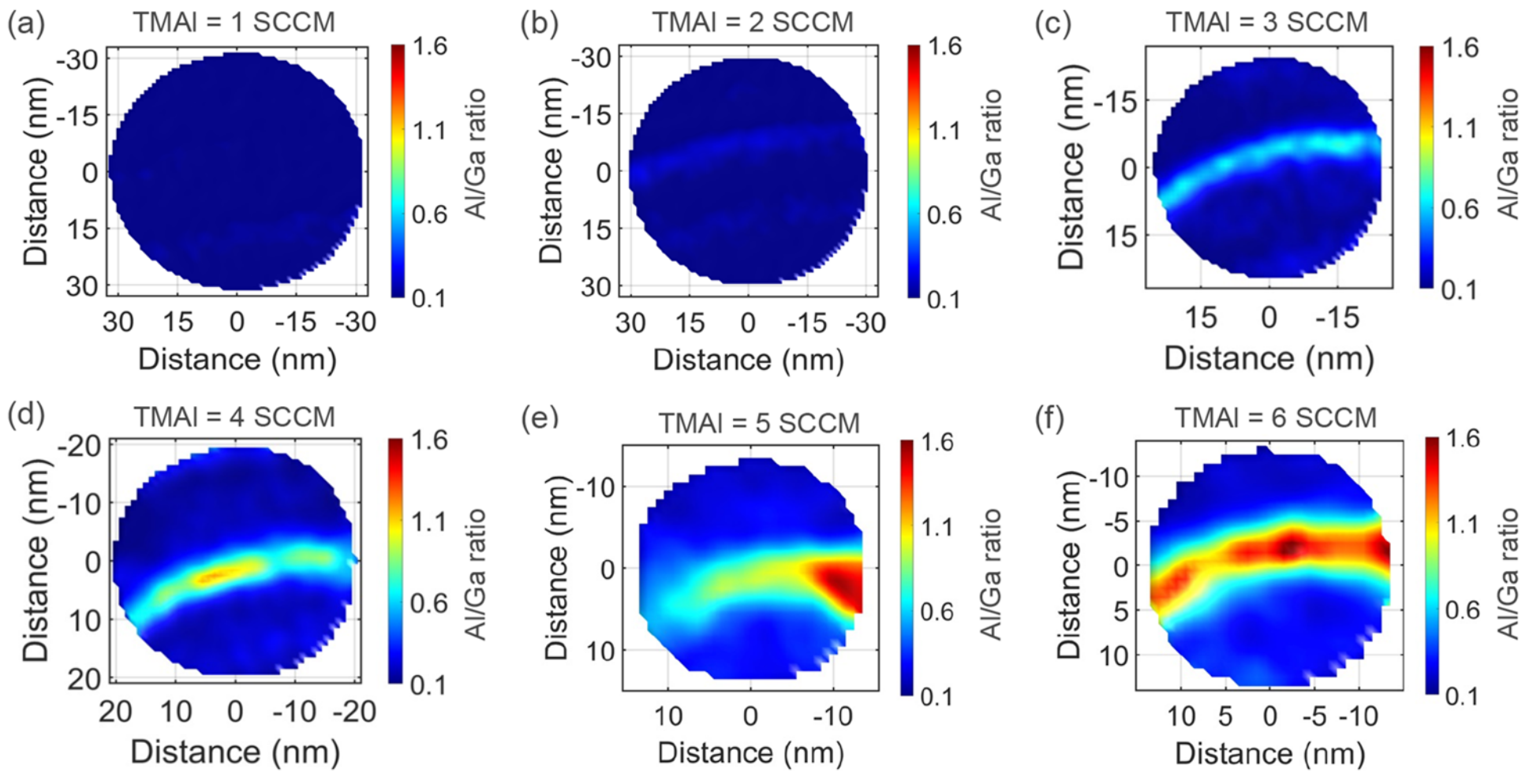

FIG. 11. (a)–(f) In-plane Al/Ga distributions for each (001) $\beta$-$(Al_xGa_{1-x})_2O_3$ layer grown with TMAl flow rates of 1–6 SCCM, respectively.

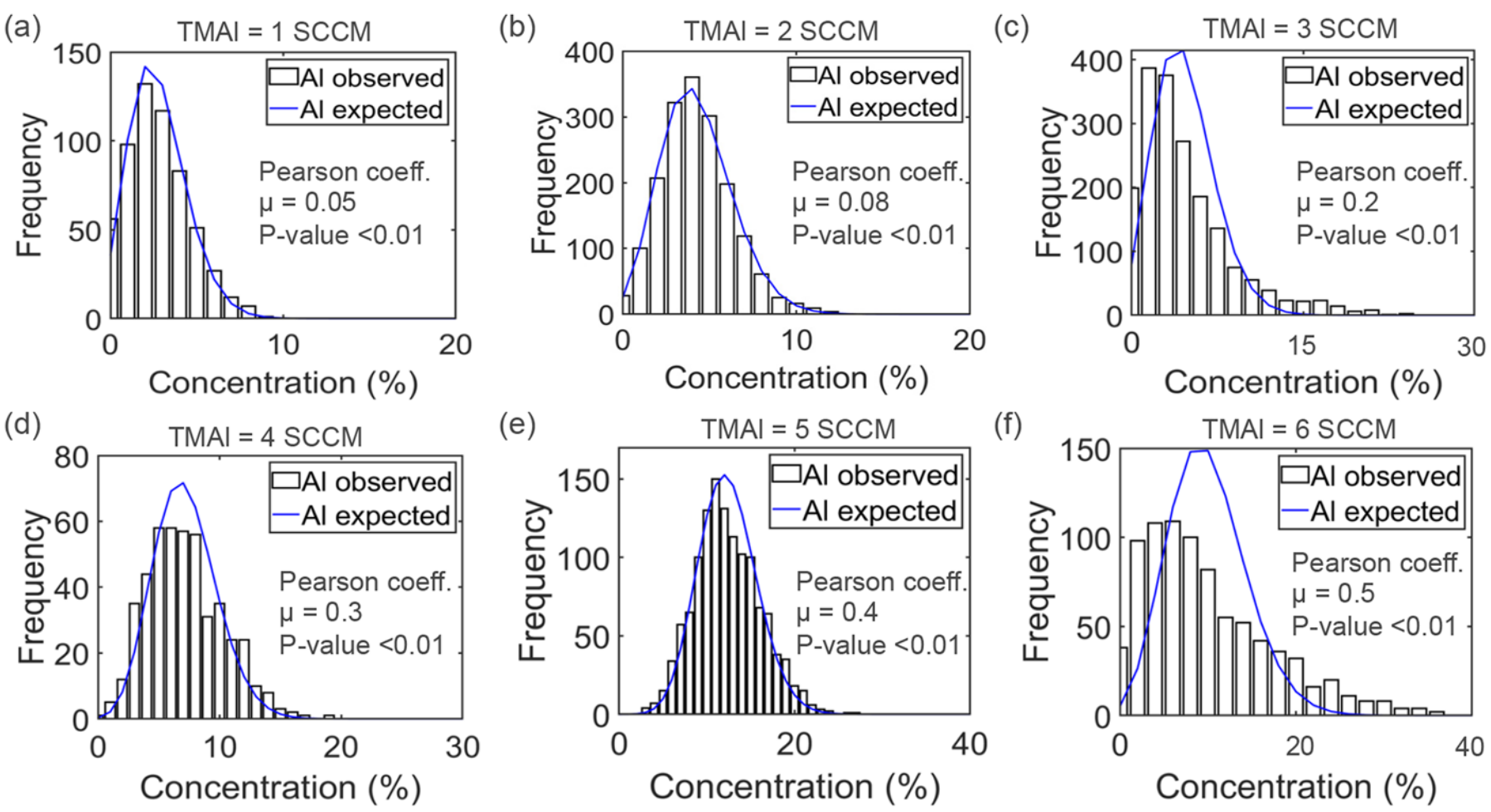

**FIG. 12.** (a)–(f) Frequency distribution analysis of Al in $\beta$-$(Al_xGa_{1-x})_2O_3$ layers grown with TMAl flow rates of 1–6 SCCM, respectively.

In addition to the investigation of the structural quality and chemical homogeneity, the band gaps of $\beta$-$(Al_xGa_{1-x})_2O_3$ thin films and the band offsets at $\beta$-$(Al_xGa_{1-x})_2O_3$/$\beta$-$Ga_2O_3$ heterojunctions were also evaluated by using high-resolution XPS. The determination of the bandgap energy using the XPS technique by examining the onset of the inelastic spectrum at the higher binding energy side of a strong intensity core level peak position is considered as a well-known method.[42–45] For the bandgap determination, Ga $2p_{3/2}$ core levels of both $\beta$-$(Al_xGa_{1-x})_2O_3$ and $\beta$-$Ga_2O_3$ thin films were used, as exemplarily shown in Fig. 13(a) for $\beta$-$(Al_xGa_{1-x})_2O_3$ films with 25% Al composition. The inset of the figure represents the zoomed-in view of the background region of Ga $2p_{3/2}$ core level. The intersection of the linear extrapolation of the loss spectra curve and the constant background is used to identify the onset of the inelastic loss spectra. From the difference between the onset of inelastic background and peak position of the Ga $2p_{3/2}$ core level, the bandgaps of 4.88, 5.04, and 5.15 eV are determined for $\beta$-$Ga_2O_3$ and $\beta$-$(Al_xGa_{1-x})_2O_3$ films with 11% and 25% Al compositions, respectively. The theoretically predicted bandgap values based on first-principles hybrid density functional theory (DFT)[1] matches well with the experimental bandgap energies obtained in this study by utilizing XPS as shown in Fig. 13(b).

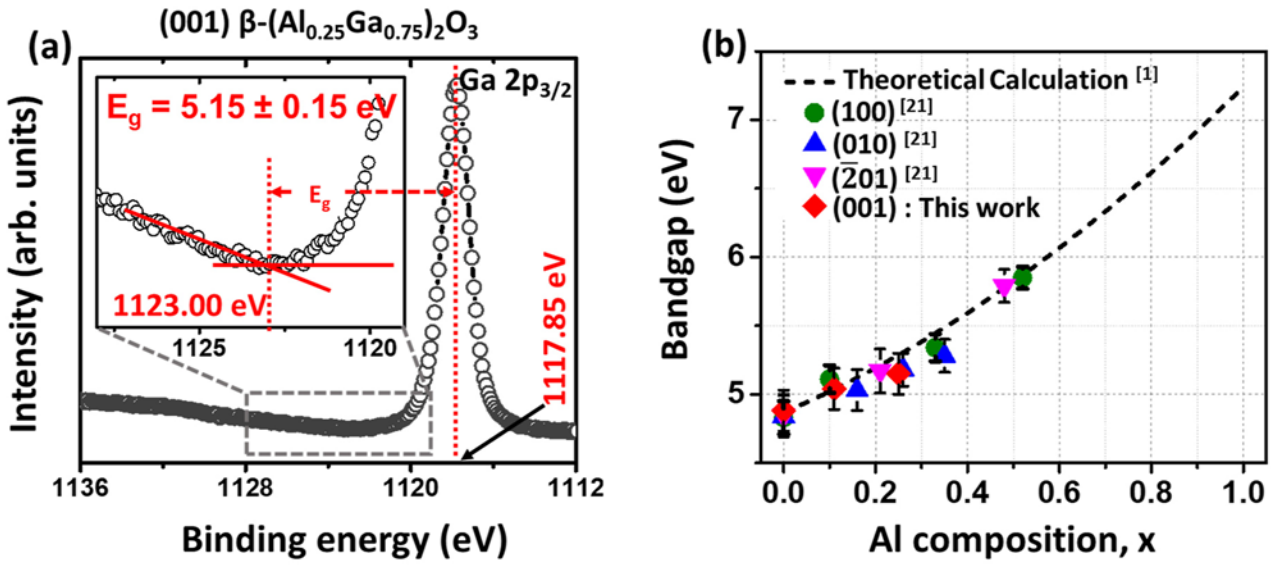

**FIG. 13.** (a) Ga $2p_{3/2}$ core level spectra of (001) oriented $\beta$-$(Al_xGa_{1-x})_2O_3$ film for x = 25% with the peak position located at 1117.85 eV binding energy. The inset shows the zoomed-in view of the inelastic background region. The bandgap of 5.15 ± 0.15 eV is calculated from the difference between the position of Ga $2p_{3/2}$ core-level and the onset of the inelastic background. (b) Bandgap energies of $\beta$-$(Al_xGa_{1-x})_2O_3$ films grown on differently oriented $\beta$-$Ga_2O_3$ substrates as a function of Al composition.

The band offsets at $\beta$-$(Al_xGa_{1-x})_2O_3$/$\beta$-$Ga_2O_3$ heterointerfaces are also determined using XPS for the Al compositions of 11% and 25%. Three types of samples were prepared for the band offset measurement: (a) 50 nm thick $\beta$-$Ga_2O_3$ film, (b) 200 nm thick $\beta$-$(Al_xGa_{1-x})_2O_3$ layers with 11% and 25% Al compositions, and (c) thin (2 nm) layer of $\beta$-$(Al_xGa_{1-x})_2O_3$ with x = 11% and 25% grown on 65 nm thick (001) $\beta$-$Ga_2O_3$ film for capturing all the electronic states from the heterointerfaces. The Kraut's method is employed for the determination of the valence ($\Delta E_v$) band offsets as follows:[46]

$$\Delta E_v = (E^{GaO}_{Ga3d} - E^{GaO}_{VBM}) - (E^{AlGaO}_{Al2p} - E^{AlGaO}_{VBM}) - (E^{AlGaO/GaO}_{Ga3d} - E^{AlGaO/GaO}_{Al2p}) \quad (1)$$

Using the extracted valence band offsets and bandgaps of $\beta$-$Ga_2O_3$ and $\beta$-$(Al_xGa_{1-x})_2O_3$ layers, the conduction band offsets ($\Delta E_c$) at the $\beta$-$(Al_xGa_{1-x})_2O_3$/$\beta$-$Ga_2O_3$ heterointerfaces are determined as follows:

$$\Delta E_c = E^{AlGaO}_g - E^{GaO}_g - \Delta E_v \quad (2)$$

Here, the binding energy corresponding to the valence band minimum (VBM) and Ga 3d core levels of 50 nm thick $\beta$-$Ga_2O_3$ films are defined as $E^{GaO}_{VBM}$ and $E^{GaO}_{Ga3d}$, respectively. Similarly, $E^{AlGaO}_{VBM}$ and $E^{AlGaO}_{Al2p}$ represent the VBM and the Al 2p core levels of 200 nm thick β-$(Al_xGa_{1-x})_2O_3$ layers. The Ga 3d and Al 2p core levels of β-$(Al_xGa_{1-x})_2O_3$/β-$Ga_2O_3$ heterointerfaces are represented by $E^{AlGaO/GaO}_{Ga3d}$ and $E^{AlGaO/GaO}_{Al2p}$, respectively. The bandgaps of β-$Ga_2O_3$ and β-$(Al_xGa_{1-x})_2O_3$ are defined as $E^{GaO}_g$ and $E^{AlGaO}_g$, respectively.

Figures 14(a)–14(c) show Ga 3d, Al 2p core levels and VBM of β-$Ga_2O_3$ and β-$(Al_xGa_{1-x})_2O_3$ layers as well as β-$(Al_xGa_{1-x})_2O_3$/β-$Ga_2O_3$ heterojunctions for 25% Al composition. After applying the Shirley background subtraction, the core-level peak positions are calculated by fitting with Gaussian and Lorentzian line shapes. The valence band onsets are calculated by linearly extrapolating the leading edge to the background. Using Eq. (1), the $\Delta E_v$ of −0.04 and −0.08 eV are determined at β-$(Al_xGa_{1-x})_2O_3$/β-$Ga_2O_3$ interfaces with Al compositions of 11% and 25%, respectively. The corresponding $\Delta E_c$ of 0.17 and 0.35 eV are calculated using Eq. (2). The measured core-level binding energies, band gaps, valence, and conduction band offsets for different Al compositions are summarized in Table I. The heterointerfaces for both Al compositions show type-II (staggered) band alignment, which is consistent with the theoretical DFT predictions for (001) orientation.[47] Previously, similar type-II band alignments were also observed at (010) and (100) β-$(Al_xGa_{1-x})_2O_3$/β-$Ga_2O_3$ interfaces, whereas $(\bar{2}01)$ oriented β-$(Al_xGa_{1-x})_2O_3$/β-$Ga_2O_3$ exhibited type-I (straddling) band alignment.[21]

The determined conduction band offsets at the interfaces between (001) β-$Ga_2O_3$ and β-$(Al_xGa_{1-x})_2O_3$ as a function of Al compositions are shown in Fig. 15. The evolution of both experimentally and theoretically determined $\Delta E_c$ values as a function of Al compositions for differently oriented β-$(Al_xGa_{1-x})_2O_3$ films, such as (010), (100), and $(\bar{2}01)$ are also included in the figure.[20,21] The conduction band offsets exhibit a strong orientation dependence. For all orientations, the conduction band offsets are found to increase as the Al composition increases. However, among all the orientations, the (100) oriented β-$(Al_xGa_{1-x})_2O_3$/β-$Ga_2O_3$ interfaces show highest conduction band offsets, which is in a good agreement with the theoretical DFT predictions.[47]

Finally, the n-type conductivity of (001) β-$(Al_xGa_{1-x})_2O_3$ thin films for different Al compositions are also investigated by using Si as a dopant. Figure 16 shows the room temperature Hall mobility vs carrier concentration of Si-doped (001) β-$(Al_xGa_{1-x})_2O_3$ films for different Al compositions. 200 nm thick β-$(Al_xGa_{1-x})_2O_3$ films with 11% Al compositions were grown with 0.53 nmole/min of silane flow rate and β-$(Al_xGa_{1-x})_2O_3$ films with 14% and 25% Al compositions were grown with the silane flow of 3.1 nmole/min. All the films exhibit n-type conductivity with room temperature Hall mobility of 39–70 $cm^2$/Vs and carrier concentrations ranging between 2 and $4 \times 10^{17}$ $cm^{-3}$. The results show decent electrical properties of (001) oriented β-$(Al_xGa_{1-x})_2O_3$ films for all the investigated Al compositions. We observed a reduction of the carrier concentration as the

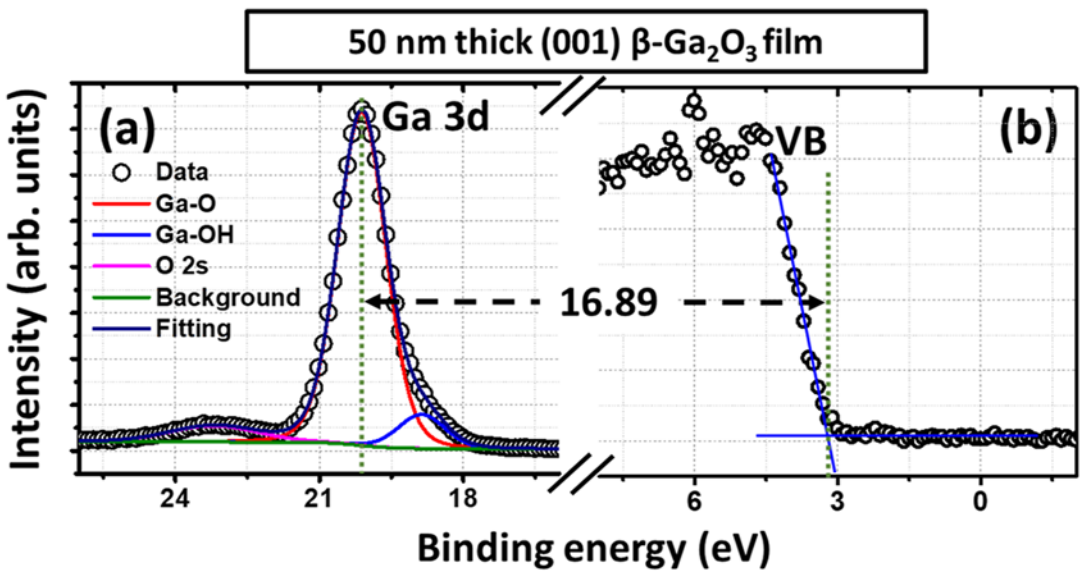

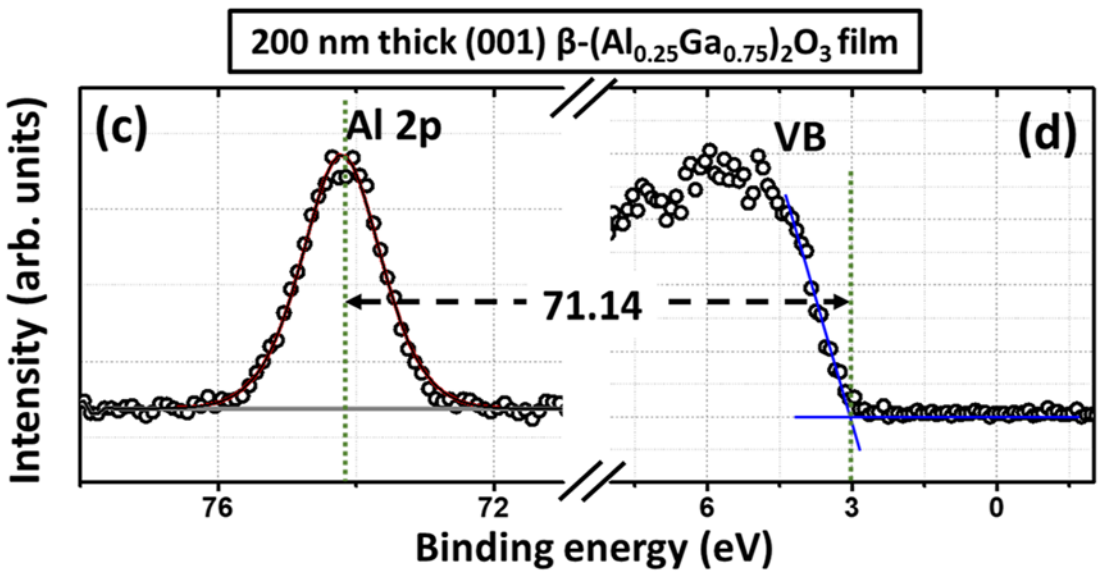

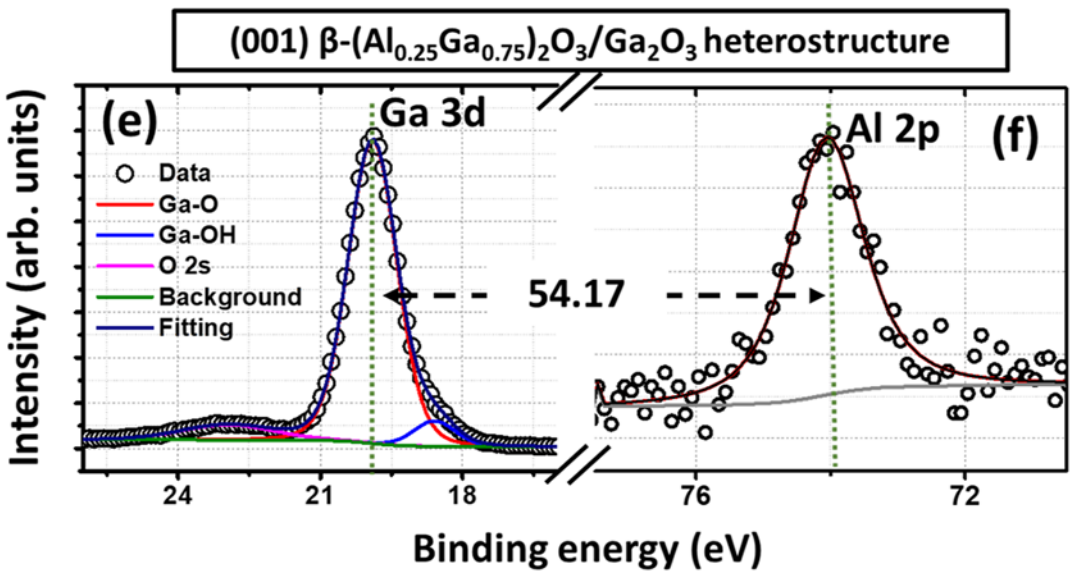

**FIG. 14.** Ga 3d and Al 2p core-levels and valence band (VB) spectra from (a) 50 nm thick (001) β-$Ga_2O_3$ film, (b) 200 nm thick β-$(Al_xGa_{1-x})_2O_3$ film, and (c) β-$(Al_xGa_{1-x})_2O_3$/β-$Ga_2O_3$ (2/65 nm) interface with x = 25%. Experimental data points are shown as black open circles, and the fitted curves are represented as black dashed lines. The navy blue solid straight lines in (b) and (d) represent the linear fitting of VB spectra of β-$Ga_2O_3$ and β-$(Al_xGa_{1-x})_2O_3$ films, respectively.

**TABLE I.** Summary of the valence and conduction band offsets at (001) β-$(Al_xGa_{1-x})_2O_3$/β-$Ga_2O_3$ interfaces, estimated by using valence band maximum, Ga 3d and Al 2p core levels, and VBM positions from the XPS measurement.

| Al composition (%) | Bandgap energy (eV) (±0.15 eV) | $(E^{GaO}_{Ga\ 3d} - E^{GaO}_{VBM})$ (eV) (±0.04 eV) | $(E^{AlGaO}_{Al\ 2p} - E^{AlGaO}_{VBM})$ (eV) (±0.04 eV) | $(E^{AlGaO/GaO}_{Ga\ 3d} - E^{AlGaO/GaO}_{Al\ 2p})$ (eV) (±0.02 eV) | $\Delta E_v$ (eV) (±0.06 eV) | $\Delta E_c$ (eV) (±0.22 eV) |
|---|---|---|---|---|---|---|
| 0 | 4.88 | 16.89 | | | | |
| 11 | 5.04 | | 71.00 | −54.07 | −0.04 | 0.17 |
| 25 | 5.15 | | 71.14 | −54.17 | −0.08 | 0.35 |

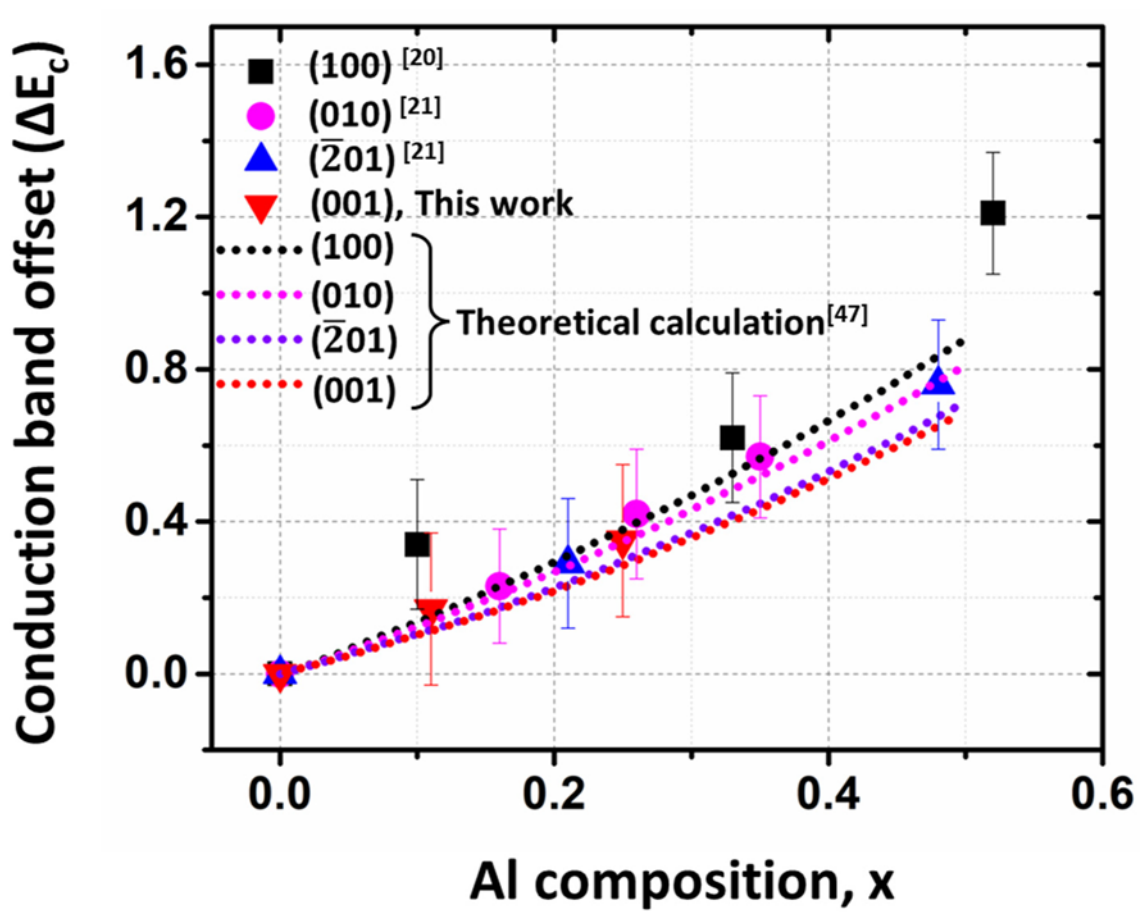

FIG. 15. Conduction band offsets at β-$(Al_xGa_{1-x})_2O_3$/β-$Ga_2O_3$ interfaces as a function of Al composition for (010), (100), ($\bar{2}$01), and (001) orientations. The dotted lines represent the theoretical predictions of the conduction band offsets for different orientations.

Al composition increases from 14% to 25% with the same silane flow rate of 3.1 nmole/min. This is expected due to the widening of the bandgap energy with the increase of Al composition, leading to higher Si donor activation energy in β-$(Al_xGa_{1-x})_2O_3$. In addition, the carrier concentration can also be reduced due to the compensation by a higher level of native defects in a high Al composition film. Our previous study on (010) β-$(Al_xGa_{1-x})_2O_3$ showed that the unintentional incorporation of C and H impurities, which act as the compensating acceptors in β-$(Al_xGa_{1-x})_2O_3$, increase with the increase in Al composition, leading to a lower net carrier concentration.[10] The reduction of the Hall mobility with increasing Al compositions from 14% to 25% can be attributed to the alloy scattering and the increase of defect concentration. However, further investigation beyond the scope of this work is necessary to fully understand the n-type doping in (001) β-$(Al_xGa_{1-x})_2O_3$.

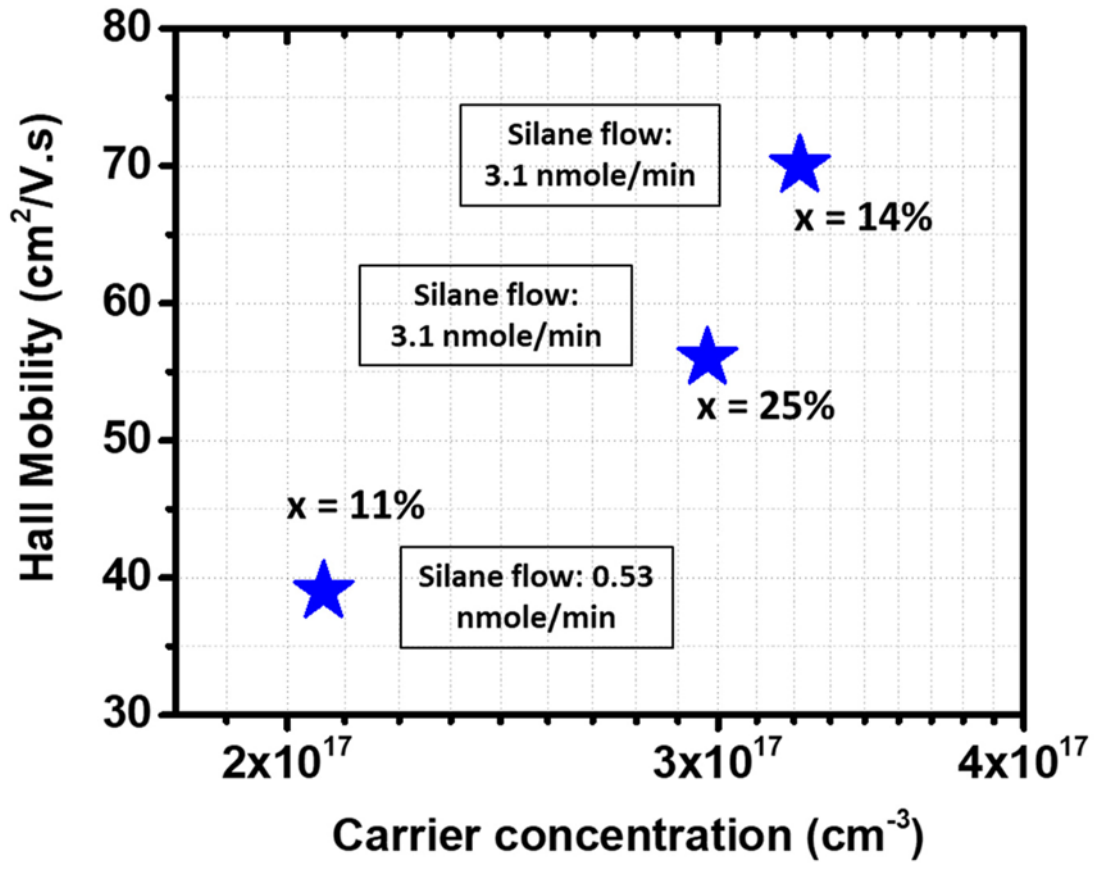

FIG. 16. Room temperature Hall mobility vs carrier concentration for β-$(Al_xGa_{1-x})_2O_3$ films grown on (001) oriented β-$(Al_xGa_{1-x})_2O_3$ substrates with various Al compositions.

## IV. CONCLUSION

In summary, the epitaxial growth of β-$(Al_xGa_{1-x})_2O_3$ thin films and β-$(Al_xGa_{1-x})_2O_3$/β-$Ga_2O_3$ superlattice structures on (001) oriented β-$Ga_2O_3$ substrates is comprehensively investigated by MOCVD. The structural, chemical, and surface morphological properties of β-$(Al_xGa_{1-x})_2O_3$ films with different Al compositions, including Al incorporation, strain, surface rms roughness, alloy homogeneity, bandgap, and band offsets are investigated by extensive material characterizations. XRD, STEM EDX elemental mapping, and APT measurements show up to 25% Al incorporation in coherently strained β-$(Al_xGa_{1-x})_2O_3$ thin films grown along the (001) orientation. However, chemical inhomogeneity with directional dependence of Al distribution along the ($\bar{2}$01) primary cleavage plane is observed in the films. The determined band offsets at β-$(Al_xGa_{1-x})_2O_3$/β-$Ga_2O_3$ interfaces with different Al compositions reveal the formation of type-II band alignment. The n-type conductivity is also investigated for different Al compositions. The results from this study on the epitaxial growth of (001) oriented β-$(Al_xGa_{1-x})_2O_3$ alloys are promising for the development of future high-power and high-frequency electronic and optoelectronic devices based on the β-$(Al_xGa_{1-x})_2O_3$/β-$Ga_2O_3$ heterostructure.

## ACKNOWLEDGMENTS

The authors acknowledge the Air Force Office of Scientific Research Grant No. FA9550-18-1-0479 (AFOSR, Dr. Ali Sayir) for financial support. The electron microscopy was performed at the Center for Electron Microscopy and Analysis (CEMAS) at The Ohio State University. The authors also acknowledge the National Science Foundation (Grant Nos. 2231026 and 2019753) and Semiconductor Research Corporation (SRC) under the Task ID No. GRC 3007.001 for partial support.

## AUTHOR DECLARATIONS

### Conflict of Interest

The authors have no conflicts to disclose.

### Author Contributions

**A. F. M. Anhar Uddin Bhuiyan**: Conceptualization (equal); Data curation (equal); Formal analysis (equal); Investigation (lead); Methodology (equal); Validation (equal); Visualization (equal); Writing – original draft (lead); Writing – review & editing (equal). **Lingyu Meng**: Formal analysis (equal); Investigation (supporting); Writing – review & editing (equal). **Hsien-Lien Huang**: Formal analysis (equal); Investigation (equal); Validation (equal); Visualization (equal); Writing – original draft (supporting); Writing – review & editing (supporting). **Jith Sarker**: Formal analysis (equal); Investigation (equal); Writing – original draft (supporting); Writing – review & editing (equal). **Chris Chae**: Investigation (supporting); Validation (supporting). **Baishakhi Mazumder**: Formal analysis (equal); Investigation (equal); Supervision (equal); Validation (equal); Writing – review & editing (equal). **Jinwoo Hwang**: Funding acquisition (equal); Investigation (equal); Methodology (equal); Supervision (equal); Validation (equal); Writing – original draft (equal). **Hongping Zhao**: Conceptualization (equal); Formal analysis (equal); Funding acquisition (equal); Investigation (equal); Methodology (equal); Project administration (equal); Resources (equal); Supervision (equal); Validation (equal); Writing – original draft (supporting); Writing – review & editing (equal).

## DATA AVAILABILITY

The data that support the findings of this study are available from the corresponding author upon reasonable request.

## REFERENCES

[1] H. Peelaers, J. B. Varley, J. S. Speck, and C. G. Van de Walle, Appl. Phys. Lett. **112**, 242101 (2018).
[2] J. B. Varley, J. Mater. Res. **36**, 4790 (2021).
[3] J. B. Varley, A. Perron, V. Lordi, D. Wickramaratne, and J. L. Lyons, Appl. Phys. Lett. **116**, 172104 (2020).
[4] K. Ghosh and U. Singisetti, J. Mater. Res. **32**, 4142 (2017).
[5] S. Krishnamoorthy, Z. Xia, C. Joishi, Y. Zhang, J. McGlone, J. Johnson, M. Brenner, A. R. Arehart, J. Hwang, S. Lodha, and S. Rajan, Appl. Phys. Lett. **111**, 023502 (2017).
[6] A. Kumar, K. Ghosh, and U. Singisetti, J. Appl. Phys. **128**, 105703 (2020).
[7] Y. Zhang, A. Neal, Z. Xia, C. Joishi, J. M. Johnson, Y. Zheng, S. Bajaj, M. Brenner, D. Dorsey, K. Chabak, G. Jessen, J. Hwang, S. Mou, J. P. Heremans, and S. Rajan, Appl. Phys. Lett. **112**, 173502 (2018).
[8] P. Ranga, A. Bhattacharyya, A. Chmielewski, S. Roy, R. Sun, M. A. Scarpulla, N. Alem, and S. Krishnamoorthy, Appl. Phys. Express **14**, 025501 (2021).
[9] A. F. M. Anhar Uddin Bhuiyan, Z. Feng, J. M. Johnson, Z. Chen, H.-L. Huang, J. Hwang, and H. Zhao, Appl. Phys. Lett. **115**, 120602 (2019).
[10] A. F. M. A. U. Bhuiyan, Z. Feng, L. Meng, A. Fiedler, H.-L. Huang, A. T. Neal, E. Steinbrunner, S. Mou, J. Hwang, S. Rajan, and H. Zhao, J. Appl. Phys. **131**, 145301 (2022).
[11] A. F. M. A. U. Bhuiyan, Z. Feng, L. Meng, and H. Zhao, J. Mater. Res. **36**, 4804 (2021).
[12] S. W. Kaun, F. Wu, and J. S. Speck, J. Vac. Sci. Technol. A **33**, 041508 (2015).
[13] A. F. M. A. U. Bhuiyan, Z. Feng, J. M. Johnson, H.-L. Huang, J. Sarker, M. Zhu, M. R. Karim, B. Mazumder, J. Hwang, and H. Zhao, APL Mater. **8**, 031104 (2020).
[14] J. M. Johnson, H.-L. Huang, M. Wang, S. Mu, J. B. Varley, A. F. M. A. Uddin Bhuiyan, Z. Feng, N. K. Kalarickal, S. Rajan, H. Zhao, C. G. Van de Walle, and J. Hwang, APL Mater. **9**, 051103 (2021).
[15] A. F. M. A. U. Bhuiyan, Z. Feng, J. M. Johnson, H.-L. Huang, J. Sarker, M. Zhu, M. R. Karim, B. Mazumder, J. Hwang, and H. Zhao, APL Mater. **8**(8), 089102 (2020).
[16] J. Sarker, A. F. M. A. U. Bhuiyan, Z. Feng, H. Zhao, and B. Mazumder, J. Phys. D: Appl. Phys. **54**, 184001 (2021).
[17] J. Sarker, S. Broderick, A. F. M. A. U. Bhuiyan, Z. Feng, H. Zhao, and B. Mazumder, Appl. Phys. Lett. **116**, 152101 (2020).
[18] A. F. M. Anhar Uddin Bhuiyan, Z. Feng, J. M. Johnson, H.-L. Huang, J. Hwang, and H. Zhao, Cryst. Growth Des. **20**, 6722 (2020).
[19] A. F. M. A. U. Bhuiyan, Z. Feng, J. M. Johnson, H.-L. Huang, J. Hwang, and H. Zhao, Appl. Phys. Lett. **117**, 142107 (2020).
[20] A. F. M. A. U. Bhuiyan, Z. Feng, J. M. Johnson, H.-L. Huang, J. Hwang, and H. Zhao, Appl. Phys. Lett. **117**, 252105 (2020).
[21] A. F. M. A. U. Bhuiyan, Z. Feng, H.-L. Huang, L. Meng, J. Hwang, and H. Zhao, J. Vac. Sci. Technol. A **39**, 063207 (2021).
[22] H. Murakami, K. Nomura, K. Goto, K. Sasaki, Q. T. Kawara, K. Thieu, R. Togashi, Y. Kumagai, M. Higashiwaki, A. Kuramata, S. Yamakoshi, B. Monemar, and A. Koukitu, Appl. Phys. Express **8**, 015503 (2015).
[23] W. Li, K. Nomoto, Z. Hu, D. Jena, and H. G. Xing, IEEE Electron Device Lett. **41**, 107–110 (2020).
[24] K. Konishi, K. Goto, H. Murakami, Y. Kumagai, A. Kuramata, S. Yamakoshi, and M. Higashiwaki, Appl. Phys. Lett. **110**, 103506 (2017).
[25] K. Sasaki, D. Wakimoto, Q. T. Thieu, Y. Koishikawa, A. Kuramata, M. Higashiwaki, and S. Yamakoshi, IEEE Electron Device Lett. **38**, 783 (2017).
[26] Q. Yan, H. Gong, J. Zhang, J. Ye, H. Zhou, Z. Liu, S. Xu, C. Wang, Z. Hu, Q. Feng, J. Ning, C. Zhang, P. Ma, R. Zhang, and Y. Hao, Appl. Phys. Lett. **118**, 122102 (2021).
[27] W. Li, K. Nomoto, Z. Hu, N. Tanen, K. Sasaki, A. Kuramata, D. Jena, and H. G. Xing, "*1.5 kV vertical* $Ga_2O_3$ *trench-MIS Schottky barrier diodes*," in *2018 76th Device Research Conference (DRC)* (IEEE, 2018), pp. 1–2.
[28] A. Mauze, T. Itoh, Y. Zhang, E. Deagueros, F. Wu, and J. S. Speck, J. Appl. Phys. **132**, 115302 (2022).
[29] M. K. Miller, K. F. Russell, K. Thompson, R. Alvis, and D. J. Larson, Microsc. Microanal. **13**, 428 (2007).
[30] T. Oshima, T. Okuno, N. Arai, Y. Kobayashi, and S. Fujita, Jpn. J. Appl. Phys. **48**, 070202 (2009).
[31] T. Shibata, H. Irie, and K. Hashimoto, J. Phys. Chem. B **107**(39), 10696 (2003).
[32] Y. Oshima, E. Ahmadi, S. C. Badescu, F. Wu, and J. S. Speck, Appl. Phys. Express **9**, 061102 (2016).
[33] K. Kaneko, K. Suzuki, Y. Ito, and S. Fujita, J. Cryst. Growth **436**, 150 (2016).
[34] F. A. Cotton, *Chemical Applications of Group Theory*, 2nd ed. (Wiley-Interscience, New York, 1971).
[35] T. Onuma, S. Fujioka, T. Yamaguchi, Y. Itoh, M. Higashiwaki, K. Sasaki, T. Masui, and T. Honda, J. Cryst. Growth **401**, 330 (2014).
[36] C. Kranert, C. Sturm, R. Schmidt-Grund, and M. Grundmann, Sci. Rep. **6**, 35964 (2016).
[37] D. Dohy, G. Lucazeau, and A. Revcolevschi, J. Solid State Chem. **45**, 180 (1982).
[38] S. Mu, M. Wang, H. Peelaers, and C. G. Van de Walle, APL Mater. **8**, 091105 (2020).
[39] B. Mazumder, J. Sarker, Y. Zhang, J. M. Johnson, M. Zhu, S. Rajan, and J. Hwang, Appl. Phys. Lett. **115**, 132105 (2019).
[40] Y. R. Luo, "Bond dissociation energies," in *CRC Handbook of Chemistry and Physics*, 90th ed., edited by D. R. Lide (CRC Press/Taylor and Francis, Boca Raton, FL, 2009).
[41] A. Devaraj, M. Gu, R. Colby, P. Yan, C. M. Wang, J. M. Zheng, J. Xiao, A. Genc, J. G. Zhang, I. Belharouak, D. Wang, K. Amine, and S. Thevuthasan, Nat. Commun. **6**, 8014 (2015).
[42] M. T. Nichols, W. Li, D. Pei, G. A. Antonelli, Q. Lin, S. Banna, Y. Nishi, and J. L. Shohet, J. Appl. Phys. **115**, 094105 (2014).
[43] A. F. M. A. U. Bhuiyan, Z. Feng, H.-L. Huang, L. Meng, J. Hwang, and H. Zhao, APL Mater. **9**, 101109 (2021).
[44] A. F. M. A. U. Bhuiyan, Z. Feng, H.-L. Huang, L. Meng, J. Hwang, and H. Zhao, J. Vac. Sci. Technol. A **40**, 062704 (2022).
[45] A. F. M. A. U. Bhuiyan, L. Meng, H.-L. Huang, J. Hwang, and H. Zhao, J. Appl. Phys. **132**, 165301 (2022).
[46] E. A. Kraut, R. W. Grant, J. R. Waldrop, and S. P. Kowalczyk, Phys. Rev. Lett. **44**, 1620 (1980).
[47] S. Mu, H. Peelaers, Y. Zhang, M. Wang, and C. G. Van de Walle, Appl. Phys. Lett. **117**, 252104 (2020).